\documentclass[twocolumn]{aastex631}
\usepackage{graphicx}	
\usepackage[caption=false]{subfig}
\usepackage{rotating}

\usepackage{amssymb}
\usepackage{color}
\usepackage{amsmath}
\usepackage{longtable}
\usepackage{threeparttablex}

\usepackage{natbib}
\newcommand{\ie}{\textit{i.e.,}}
\newcommand{\eg}{\textit{e.g.,}}

\newcommand{\logM}{$\log(M_*/\rm{M}_\odot)$}
\newcommand{\zphot}{$z_\textrm{phot}$}

\newcommand{\pz}{$p(z)$}

\newcommand{\OII}{\hbox{{\rm [O}\kern 0.1em{\sc ii}{\rm ]$\lambda\lambda3726,3729$}}}
\newcommand{\OIIIfive}{\hbox{{\rm [O}\kern 0.1em{\sc iii}{\rm ]$\lambda5007$}}}
\newcommand{\OIIIfour}{\hbox{{\rm [O}\kern 0.1em{\sc iii}{\rm ]$\lambda4959$}}}
\newcommand{\OIII}{\hbox{{\rm [O}\kern 0.1em{\sc iii}{\rm ]}}}

\newcommand{\Hbeta}{$\rm{H}\beta$}
\newcommand{\Halpha}{$\rm{H}\alpha$}
\newcommand{\NII}{[N~{\rm \scriptsize II}]$\lambda\lambda6548,6584$}
\newcommand{\SII}{[S~{\rm \scriptsize II}]$\lambda\lambda6718,6733$}

\newcommand{\dv}{$\Sigma_{{\rm VMC},\alpha,\delta,z}$}
\newcommand{\mdv}{$\tilde{\Sigma}_{{\rm VMC},z}$}
\newcommand{\odv}{$\log(1+\delta_{{\rm VMC},\alpha,\delta,z})$}
\newcommand{\odf}{$\log(1+\delta_{f^5,\alpha,\delta,z})$}
\newcommand{\od}{$\log(1+\delta_{\rm gal})$}

\newcommand{\specsamp}{SSR}
\newcommand{\szf}{S$z$F}
\newcommand{\PSC}{\mbox{Elent\'ari}}

\begin{document}

\title{\sc Environmental Effects on the Stellar Mass Function in a $z\sim3.3$ Overdensity of Galaxies in the COSMOS Field\footnote{Some of the data presented herein were obtained at Keck Observatory, which is a private 501(c)3 non-profit organization operated as a scientific partnership among the California Institute of Technology, the University of California, and the National Aeronautics and Space Administration. The Observatory was made possible by the generous financial support of the W. M. Keck Foundation.}}
\shorttitle{$z\sim3.3$ Environmental Stellar Mass Functions}
\shortauthors{B. Forrest, et al.}

\correspondingauthor{Ben Forrest}
\email{bforrest@ucdavis.edu}

\author[0000-0001-6003-0541]{Ben Forrest}
	\affiliation{Department of Physics and Astronomy, University of California Davis, One Shields Avenue, Davis, CA, 95616, USA}
\author[0000-0002-1428-7036]{Brian C. Lemaux}
	\affiliation{Gemini Observatory, NSF's NOIRLab, 670 N. A'ohoku Place, Hilo, HI, 96720, USA}
	\affiliation{Department of Physics and Astronomy, University of California Davis, One Shields Avenue, Davis, CA, 95616, USA}
\author[0000-0001-7811-9042]{Ekta A. Shah}
	\affiliation{Department of Physics and Astronomy, University of California Davis, One Shields Avenue, Davis, CA, 95616, USA}
\author[0000-0002-8877-4320]{Priti Staab}	
	\affiliation{Department of Physics and Astronomy, University of California Davis, One Shields Avenue, Davis, CA, 95616, USA}
\author[0000-0001-8255-6560]{Roy R. Gal}
	\affiliation{University of Hawai'i, Institute for Astronomy, 2680 Woodlawn Drive, Honolulu, HI 96822, USA}
\author[0000-0003-2119-8151]{Lori M. Lubin}
	\affiliation{Department of Physics and Astronomy, University of California Davis, One Shields Avenue, Davis, CA, 95616, USA}
	
\author[0000-0003-1371-6019]{M. C. Cooper}
	\affiliation{Department of Physics and Astronomy, University of California, Irvine, 4129 Frederick Reines Hall, Irvine, CA 92697, USA}
\author[0000-0002-9336-7551]{Olga Cucciati}
	\affiliation{INAF Osservatorio di Astrofisica e Scienza dello Spazio di Bologna, Via Piero Gobetti 93/3, 40129 Bologna, Italy}
\author[0000-0001-7523-140X]{Denise Hung}
	\affiliation{Gemini Observatory, NSF's NOIRLab, 670 N. A'ohoku Place, Hilo, HI, 96720, USA}
	\affiliation{University of Hawai'i, Institute for Astronomy, 2680 Woodlawn Drive, Honolulu, HI 96822, USA}
\author[0000-0002-2446-8770]{Ian McConachie}
	\affiliation{Department of Physics \& Astronomy, University of California Riverside, 900 University Ave., Riverside, CA, 92521, USA}
\author[0000-0002-9330-9108]{Adam Muzzin}
	\affiliation{Department of Physics and Astronomy, York University, 4700, Keele Street, Toronto, ONMJ3 1P3, Canada}
\author[0000-0002-6572-7089]{Gillian Wilson}
	\affiliation{Department of Physics, University of California Merced, 5200 North Lake Rd., Merced, CA 95343, USA}
	
\author[0000-0002-8900-0298]{Sandro Bardelli}
	\affiliation{INAF Osservatorio di Astrofisica e Scienza dello Spazio di Bologna, Via Piero Gobetti 93/3, 40129 Bologna, Italy}
\author[0000-0001-5760-089X]{Letizia P. Cassar\`a}
	\affiliation{INAF-IASF Milano, Via Alfonso Corti 12, 20133, Milano, Italy}
\author[0000-0003-2144-2943]{Wenjun Chang}
	\affiliation{Department of Physics \& Astronomy, University of California Riverside, 900 University Ave., Riverside, CA, 92521, USA}
\author[0009-0003-2158-1246]{Finn Giddings}	
	\affiliation{University of Hawai'i, Institute for Astronomy, 2680 Woodlawn Drive, Honolulu, HI 96822, USA}
\author[0000-0001-5160-6713]{Emmet Golden-Marx}
	\affiliation{Department of Astronomy, Tsinghua University, Beijing 100084, People's Republic of China}
\author[0000-0001-6145-5090]{Nimish Hathi}
	\affiliation{Space Telescope Science Institute, Baltimore, MD 21218, USA}
\author[0000-0001-8169-7249]{Stephanie M. Urbano Stawinski}
	\affiliation{Department of Physics and Astronomy, University of California, Irvine, 4129 Frederick Reines Hall, Irvine, CA 92697, USA}
\author[0000-0002-5845-8132]{Elena Zucca}
	\affiliation{INAF Osservatorio di Astrofisica e Scienza dello Spazio di Bologna, Via Piero Gobetti 93/3, 40129 Bologna, Italy}


\begin{abstract}
We present an analysis of the number density of galaxies as a function of stellar mass (\ie\ the stellar mass function, SMF) in the COSMOS field at $z\sim3.3$, making a comparison between the SMF in overdense environments and the SMF in the coeval field.
In particular, this region contains the \PSC\ proto-supercluster, a system of 6 extended overdensities spanning $\sim70$~cMpc on a side.
A clear difference is seen in the high-mass slope of these SMFs, with overdense regions showing an increase in the ratio of high-mass galaxies to low-mass galaxies relative to the field, indicating a more rapid build-up of stellar mass in overdense environments.
This result qualitatively agrees with analyses of clusters at $z\sim1$, though the differences between protocluster and field SMFs at $z\sim3.3$ are smaller.
While this is consistent with overdensities enhancing the evolution of their member galaxies, potentially through increased merger rates, whether this enhancement begins in protocluster environments or even earlier in group environments is still unclear.
Though the measured fractions of quiescent galaxies between the field and overdense environments do not vary significantly, implying that this stellar mass enhancement is ongoing and any starbursts triggered by merger activity have not yet quenched, we note that spectroscopic observations are biased towards star-forming populations, particularly for low-mass galaxies.
If mergers are indeed responsible, high resolution imaging of \PSC\ and similar structures at these early epochs should then reveal increased merger rates relative to the field.
Larger samples of well-characterized overdensities are necessary to draw broader conclusions in these areas.
\end{abstract}

%%%%%%%%%%%%%%%%%%%%%%%%%%%%%%%%%%%%%%%%%%%%%%%%%%

%%%%%%%%%%%%%%%%% BODY OF PAPER %%%%%%%%%%%%%%%%%%

\section{Introduction}

The environment in which a galaxy resides plays an important role in its evolution.
Obvious differences exist in the local Universe between the populations of galaxies in massive, evolved clusters and populations of galaxies in the field across a wide range of properties including stellar mass, star formation rate, age, color, morphology, velocity dispersion, and metallicity \citep[\eg][]{Dressler1984, Kauffmann2004, Thomas2005}.
These differences are consistent with cluster galaxies having a faster and/or earlier evolutionary timescale than field galaxies and this signal is seen in the most overdense environments out past $z\gtrsim1$ \citep[\eg][]{Kawinwanichakij2017a, Papovich2018, Mei2023}.
The strength of this signal begins to decrease or perhaps reverse by $z\gtrsim1.5$ \citep{Tran2010, Nantais2016a, Perez-Martinez2022, Taamoli2023, Edward2023}, though some systems with elevated quenched fractions do still appear at earlier epochs \citep[\eg][]{Zavala2019, McConachie2022, Ito2023}.
In fact, at very early epochs, many galaxies in protoclusters, the progenitors of today's clusters, appear to have enhanced star formation rates (SFRs) relative to the field \citep[\eg][]{Capak2011, Hatch2011, Wang2016, Perez-Martinez2023, Staab2024}, suggesting that the effect of an overdense environment is to increase SFRs at early times.
This may subsequently deplete the molecular gas content of these galaxies, as seen in low redshift galaxies \citep[\eg][]{Fumagalli2009}, thus inhibiting future star formation and leaving the galaxy more massive and with older stellar populations than galaxies which continue forming stars.
This is consistent with findings of enhanced molecular gas reservoirs and gas fractions in clusters, protoclusters, and groups relative to the field at high redshift \citep[\eg][]{Noble2019, Jin2021, JShen2021}, a trend which also disappears around $z\sim1.5$ \citep[\eg][]{Alberts2022, Williams2022a}.
However, the specific physical mechanisms responsible for any such environmental effects to take place and whether there is a particular density threshold or timescale required for these effects is unclear.

The combined effects of these processes can be seen by analyzing the number density of galaxies as a function of their stellar mass, which is an integral of the SFR of a galaxy across cosmic time.
The shape of this stellar mass function (SMF) between nearby galaxies in clusters and in the field varies significantly, with galaxies in overdense regions showing a larger ratio of high-mass to low-mass galaxies \citep[\eg][]{Blanton2009}.
Some evidence suggests that this variation in shape may be entirely due to changes in shape and/or normalization of the SMF of red / early-type / quiescent populations, with the shape of the star-forming galaxy SMF staying relatively similar between field and cluster environments \citep[\eg][]{YPeng2010}, though this is not universally found \citep[\eg][]{Annunziatella2014, Annunziatella2016}.

Evidence is mixed as to whether or not this is the case in higher redshift overdensities.
Field surveys at $z\sim1$ have shown that the SMF varies with environmental density \citep{Bundy2006, Cooper2010, Papovich2018}.
Targeted spectroscopic observations of cluster environments at $z\sim1$ such as those in the ORELSE \citep{Lubin2009}, GCLASS \citep{Muzzin2012}, and GOGREEN \citep{Balogh2017, Balogh2020} surveys have found differences between cluster and field population SMFs as well.
The results from GCLASS \citep{vanderBurg2013} and GOGREEN \citep{vanderBurg2020} suggest that these differences are due to changes in the relative fraction of quenched galaxies, however the shapes of the quiescent SMFs in both overdensity and field are statistically the same, as are the shapes of the star-forming SMFs. 
However, results from ORELSE \citep{Tomczak2017} instead show evolution of shape in both the quiescent and star-forming SMFs with environmental overdensity.
While narrowband imaging of \Halpha\ emitters in two protoclusters at $z=2.16$ \citep{Shimakawa2018b} and $z=2.53$ \citep{Shimakawa2018a} show differences in SMF shape relative to the field, a compilation of results at $z\gtrsim2$ found a combined protocluster SMF with shape consistent with that of the field for star-forming galaxies, while the shape of quiescent galaxy SMFs between environments showed minor differences \citep{Edward2023}.

While SMF analyses of field populations $z>3$ are numerous \citep[\eg][]{Marchesini2009, Stefanon2015, Marsan2022, Weaver2023a}, similar analyses of overdense environments are lacking.
This stems from the difficulty of identification and characterization of high redshift overdensities, which is several-fold.
Photometric identification of candidate overdensities at these epochs requires deep multiband imaging, in particular in the near-infrared, in order to identify stellar-mass limited samples of faint galaxies, to infer the location of features such as the Balmer break which aid in constraining photometric redshifts, and to cover the large projected sizes of these structures \citep[$\sim10-15$ comoving Mpc, and in some cases larger;][]{Muldrew2015, Chiang2017, Cucciati2018}.
The low density contrast of these structures with the coeval field combined with significant photometric redshift uncertainties means that significant spectroscopic follow-up is also required to obtain precise galaxy redshifts, confirm these structures, and allow for accurate density mapping of the systems and their surroundings.

Similarly, simulations have had some success replicating quenched fractions and SMFs in the Universe over a range of redshifts \citep{Pillepich2018, deLucia2024}, though some disagreement in the SMF for high stellar mass galaxies at $z\gtrsim2.5$ has been noted \citep[\eg][]{Steinhardt2016, Sherman2020a}.
Cluster SMFs can also be well replicated out to $z\sim1.5$, though the quenched fractions of satellite galaxies are often severely different from observations \citep[\eg][]{Bahe2017, Kukstas2023}.
Some of this mismatch can be attributed to slightly different definitions of \eg\ star-forming vs. quiescent galaxies \citep{Donnari2021}.
Recent results from the GAEA models \citep{deLucia2024} appear to replicate the observed quenched fraction from GOGREEN clusters at $z\sim1$ \citep{vanderBurg2020}.

An assumption often made in attempting to understand the environmental effects on galaxy evolution is that the cluster satellite galaxies are drawn from same population as the field, and have simply fallen into a massive halo which proceeds to influence member galaxies.
Many works have also suggested that such influences begin even before infall into a massive cluster, in the earlier proto-cluster or group environments in what has been termed `pre-processing' \citep{Balogh2000, Fujita2004, DeLucia2012, Donnari2020}.
However, a recent analysis of the dark matter halos in simulations has suggested that the population of cluster galaxies may not in fact be drawn from the same parent population as field galaxies which would allow for intrinsic halo properties to be responsible for observed population differences between field and cluster environments \citep{Ahad2024}.

Exploring whether there are environmental effects on the stellar mass function in protocluster environments at $z>3$ can lead to increased understanding of the mechanisms in overdense environments which contribute to galaxy evolution.
To that end, in this work we build a stellar mass function based on galaxies in the COSMOS field around the \PSC\ proto-supercluster at $z\sim3.3$ \citep{Forrest2023}.
To our knowledge, this is the first time such an analysis has been performed at this early epoch and is only possible due to a wealth of deep photometric and spectroscopic data as described in Section~\ref{Sec:Data}.
We discuss the analysis methodology and results in Section~\ref{Sec:Analysis} before presenting conclusions (Section~\ref{Sec:Conc}).
Throughout this work we use the AB magnitude system \citep{Oke1983} and assume a $\Lambda$CDM cosmology with with $H_0 = 70$~km/s/Mpc, $\Omega_M = 0.3$, and $\Omega_\Lambda = 0.7$.

\section{Data} \label{Sec:Data}

\subsection{Parent Photometric Catalogs}

This work relies upon the considerable investment of observing time focused on the COSMOS field \citep{Scoville2007, Koekemoer2007}.
The ultraviolet, optical, and near-infrared imaging in this field have been compiled most recently in the COSMOS2020 catalogs \citep{Weaver2022}, which contain over 1.5 million sources observed in up to 40 bandpasses over $\sim1.5$~deg$^2$.
This includes space-based observations from \textit{GALEX} \citep[ultraviolet;][]{Zamojski2007}, \textit{HST}/ACS \citep[optical;][]{Leauthaud2007} and \textit{Spitzer}/IRAC
\citep[near-infrared][]{Ashby2013, Ashby2015, Ashby2018, Steinhardt2014}.
Ground based data includes ultraviolet CFHT/MegaCam observations \citep{Sawicki2019}, optical data from Subaru/Suprime-Cam \citep{Taniguchi2007,Taniguchi2015} and Hyper Suprime-Cam \citep{Aihara2019}, and near-infrared observations from VISTA/VIRCAM \citep{McCracken2012, Moneti2019}.

The depth of the imaging involved in construction of the COSMOS2020 catalogs as well as associated derived properties, including well-characterized photometric redshifts (\zphot) and their probability distributions (\pz), rest-frame colors, stellar masses, and SFRs are critical to this work.
Unless otherwise stated, we use the COSMOS2020 Classic catalog, and the associated properties derived using \texttt{LePhare} \citep{Arnouts1999, Ilbert2006}.
These characterizations use the same process as described in \citet{Ilbert2013}, and include galaxy templates from \citet{Polletta2007}, \citet{Bruzual2003}, and \citet{Onodera2012}, and stellar templates from \citet{Pickles1998}, with additional templates for white and brown dwarfs as well as AGN.
Allowed dust attenuation curves include Small Magellanic Cloud \citep{Prevot1984}, starburst \citep{Calzetti2000}, and starburst+2175{\rm \AA} profiles \citep{Fitzpatrick1986}, and a \citet{Chabrier2003} initial mass function is assumed.

\subsection{Spectroscopy}

\subsubsection{Field Surveys}\label{sec:spec}

The COSMOS field has also been the target of many spectroscopic surveys, several of which are included in this work.
The VIMOS Ultra-Deep Survey \citep[VUDS; ][]{LeFevre2015}, targeted $\sim$10$^4$ objects across the COSMOS, ECDFS, and VVDS-2h fields with the VIMOS instrument on ESO-VLT \citep{LeFevre2003}.
This survey preferentially selected targets for follow-up which had \zphot$\gtrsim2$ and was deep enough to reliably detect continuum for $i\sim25$ objects.

The zCOSMOS survey \citep{Lilly2007} also used the VIMOS spectrograph on the VLT and consisted of two sub-samples.
The zCOSMOS-bright subsample targeted $\sim2\times10^4$ galaxies with \mbox{$I<22.5$} and thus mainly confirmed galaxies across $0.1<z<1.2$ over the entire COSMOS ACS field.
The zCOSMOS-deep subsample targeted $\sim$10$^4$ galaxies in the central portion of the field believed to have $1.4<z<3.0$ based on their colors (Daichi, K. et al., in preparation).

The DEIMOS 10k Spectroscopic Survey \citep{Hasinger2018} similarly targeted $\sim$10$^4$ objects, using the DEIMOS instrument on the Keck II telescope \citep{Faber2003}. While the entirety of this survey was in the COSMOS field, there was no photometric redshift cut applied and thus the majority of objects with spectroscopic redshifts are at $z<2$.
Sample selection was heterogenous, with several subsamples of galaxies, including (amongst others) Spitzer/MIPS sources, high-redshift candidates, and optical counterparts of X-ray sources. 

These surveys have also used similar spectroscopic redshift quality flagging systems.
The base flag for each object is one of the following:
0 - no redshift measured,
1 - low confidence redshifts,
2, 3, 4, 9 - secure redshifts \citep[estimated $\gtrsim75\%$ reliability;][]{Lilly2007, LeFevre2013, Cassata2015}.
Each flag may also prepended with a number X, indicating that the target either has broad lines observed in the spectrum (X=1), the target is a serendipitous detection (X=2), or the target is a serendipitous detection at location of target (\ie chance alignment or merger; X=3).

\begin{table*}
	\centering
	\caption{C3VO Keck/MOSFIRE Observations in COSMOS. $H$-band exposures were 120s each, $K$-band exposures were 180s each. The slitwidth was 0.7\arcsec.}
	\label{tab:mosfire}
	\begin{tabular}{ccccrc}
		\hline
		Target ($z$) 	& Mask 	& Bandpass & Date(s) Observed & Total Exp.Time (m) & Avg. Seeing(\arcsec) \\
		\hline
		Hyperion (2.45)	& Hyperion1	& H		& 2020.03.09	& 150	& 0.71	\\
				& Hyperion2	& H		& 2020.03.09	& 112	& 0.69	\\
				& Hyperion3	& H		& 2020.03.10	& 106	& 0.66	\\
				& Hyperion4	& H		& 2020.03.11	& 104	& 0.55	\\
				& Hyperion5	& H		& 2020.03.11	& 106	& 0.55	\\
				& Hyperion6	& H		& 2020.11.30	& 118	& 0.79	\\
				& Hyperion7	& H		& 2020.11.30	& 164	& 0.86	\\
				&			&		& \hspace{0.5cm} 2021.01.06	&	&	\\
				&			&		& \hspace{0.5cm} 2021.12.25	&	&	\\
		\hline
		Elent\`ari (3.33)	& DONGOCps23n24\_1  & K  & 2021.12.26  & 102	& 0.77	\\
				& DONGOCps23n24\_2  & K  & 2021.12.26  & 102	& 0.68	\\
				& DONGOCps23n24\_3  & K  & 2022.10.15  & 48	& 0.86	\\
				& NEb\_1		& K		& 2023.02.03	& 90		& 0.86	\\
				& NEc\_1		& K		& 2023.03.30	& 60		& 0.67	\\
				& NEd\_1		& K		& 2023.03.30	& 42		& 0.81	\\
				& Bridge1		& K		& 2023.02.03	& 84		& 0.89	\\
				&			&		& \hspace{0.5cm} 2023.03.30	&	&	\\
				& Bridge2		& K		& 2023.03.31	& 72		& 0.70	\\
				& SWa\_1		& K		& 2023.02.03	& 84		& 1.00	\\
				&			&		& \hspace{0.5cm} 2023.03.30	&	&	\\
				& SWc\_1		& K		& 2023.03.31	& 48		& 0.93	\\
		\hline
	\end{tabular}
\end{table*}

\begin{table*}
	\centering
	\caption{C3VO Keck/DEIMOS Observations in COSMOS. The 600ZD grating (600 lines/mm) was used for all observations.}
	\label{tab:deimos}
	\begin{tabular}{cccccrc}
		\hline
		Target ($z$)  	& Mask 	& Filter  & Central Wavelength ({\rm \AA})  &Date(s) Observed & Total Exp.Time (m) & Avg. Seeing(\arcsec) \\
		\hline
		Taralay (4.57)	& dongN1C	& GG400	& 6500	& 2017.12.26	& 210	& 0.8	 \\
				& dongS1B	& GG400	& 6500	& 2016.12.22	& 275	& 0.8 \\
				& dongD1		& GG455	& 7200	& 2017.12.26 	& 320	& 0.94	\\
				&			&	&	& \hspace{0.5cm} 2019.02.05	&	&	\\
				&			&	&	& \hspace{0.5cm} 2019.12.24	&	&	\\
				&			&	&	& \hspace{0.5cm} 2020.12.23	&	&	\\
				& dongD2		& GG455	& 7200	& 2020.02.02	& 360	& 0.78	\\
				&			&	&	& \hspace{0.5cm} 2020.12.10	&	&	\\
				&			&	&	& \hspace{0.5cm} 2022.12.23	&	&	\\
				& dongA1		& GG455	& 7200	& 2021.01.17	& 299  	& 0.84	\\
				&			&	&	& \hspace{0.5cm} 2022.01.10	&	&	\\
				& dongA2		& GG455	& 7200	& 2022.01.10	& 240  	& 0.59	\\
				&			&	&	& \hspace{0.5cm} 2022.01.101	&	&	\\
		\hline
	PCl~J1000+0200 (2.90) & dongC2D1	& GG400	& 6500	& 2022.01.11	& 145	& 0.82	\\
			  & dongC2N1		& GG400	& 6500	& 2020.02.02	& 309	& 1.15	\\
			  & dongC2S1		& GG400	& 6500	& 2019.02.25	& 268	& 0.97	\\
				&			&	&	& \hspace{0.5cm} 2020.12.10	&	&	\\
				&			&	&	& \hspace{0.5cm} 2020.12.23	&	&	\\
				&			&	&	& \hspace{0.5cm} 2021.01.17	&	&	\\
		\hline
	\end{tabular}
\end{table*}

\subsubsection{Targeted Surveys - C3VO}

The Charting Cluster Construction with VUDS \citep{LeFevre2015} and ORELSE \citep{Lubin2009} Survey \citep[C3VO;][]{Lemaux2022} has used the Keck/DEIMOS and Keck/MOSFIRE \citep{McLean2010, McLean2012} instruments on the Keck telescopes to follow-up candidate overdensities identified in density maps constructed from a combination of spectroscopic and photometric data across the CFHTLS-D1, ECDFS, and COSMOS fields (see Sections~\ref{Sec:VMC} and \ref{Sec:Struct} for a description of these maps and candidate overdensity identification).
The survey has observed $\sim2000$ galaxies across the three fields, approximately half with MOSFIRE and half with DEIMOS.
Targeted regions included Hyperion \citep[$z=2.45$;][]{Casey2015,Cucciati2018},
PCl~J1000+0200 \citep[$z=2.90$;][]{Cucciati2014}, 
PCl~J0227-0421 \citep[$z=3.31$;][]{Lemaux2014,Shen2021}, 
Elent\'ari \citep[$z=3.33$;][]{McConachie2022, Forrest2023}, 
Smruti \citep[$z=3.47$;][]{Forrest2017, Shah2024}, and 
Taralay \citep[$z=4.57$;][]{Lemaux2018, Staab2024}.

All C3VO COSMOS observations used in this work are shown in Tables~\ref{tab:mosfire}~\&~\ref{tab:deimos}.
MOSFIRE masks targeting Hyperion as well as the first three masks targeting Elent\'ari used photometric redshifts and magnitudes from the COSMOS2015 catalog \citep{Laigle2016} for target selection.
Subsequent masks targeting \PSC\ (observed in 2023) selected objects based on \pz, stellar masses, and rest-frame colors from the COSMOS2020 Classic catalog.
DEIMOS masks prioritized targeting star-forming galaxies down to $i_{AB} < 25.3$ with photometric redshifts near that of the overdensity in question, as detailed in \citet{Lemaux2022}. 

In this work we focus on density maps constructed from data including VUDS and zCOSMOS spectroscopy, photometry from COSMOS2015 and COSMOS2020, and C3VO spectroscopic data taken prior to fall 2021 (the 2021B semester).
Spectroscopic redshifts from other masks listed in Tables~\ref{tab:mosfire} \& \ref{tab:deimos} are incorporated into subsequent pieces of the analysis, and are important for confirming redshifts of objects whose membership in Elent\'ari is inconclusive from \pz\ alone.

\subsubsection{Targeted Surveys - MAGAZ3NE}

The Massive Ancient Galaxies at $z>3$ Near-Inrared Survey \citep[MAGAZ3NE;][]{Forrest2020b} has used Keck/MOSFIRE to spectroscopically follow-up ultra-massive galaxies (UMGs; \logM$>11$ at $z>3$) and investigate their environments.
It has thus far targeted $\sim1000$ galaxies across the UltraVISTA (COSMOS), VIDEO-XMM, and VIDEO-CDFS fields.
This survey selected targets in the COSMOS field for follow-up based on observed galaxy spectral energy distributions (SEDs), \pz, stellar masses, and SFRs from the UltraVISTA DR1 and DR3 catalogs \citep[][A. Muzzin, private communication]{Muzzin2013a}.

This paper focuses on a set of six structures (S1-S6) at $z\sim3.3$ named \PSC\ \citep{Forrest2023}.
While it is unlikely that all six will collapse into a single system at $z=0$, there are multiple pairs of structures which may do so.
This system as a whole has over 100 spectroscopically-confirmed members, and the best characterized structure (S1) has an estimated $z=0$ mass of $1.3\times10^{14}$~M$_\odot$.
Two of these (S1, S4) were spectroscopically confirmed via MAGAZ3NE spectroscopy \citep{McConachie2022} as they contain UMGs, and the larger region was independently identified and then spectroscopically followed-up with C3VO.

\subsubsection{Spectroscopic Data Reduction}

Details on the data reduction process and redshift assignment for zCOSMOS \citep{Lilly2007}, VUDS \citep{LeFevre2015}, and DEIMOS 10k \citep{Hasinger2018} data can be found in the publications of those surveys. 
However, all surveys used a combination of automated initial redshift assignment followed by visual inspection and checking of results.

We reduced the C3VO MOSFIRE spectra using the MOSDEF 2D data reduction pipeline \citep{Kriek2015}.
This pipeline was also used to re-reduce the MAGAZ3NE MOSFIRE data, which were initially reduced using the Keck-supported MOSFIRE Data Reduction Pipeline (DRP; Version 2018).
The MOSDEF pipeline subtracts the sky background noise, masks both cosmic rays and bad pixels, and rectifies each frame.
It also identifies the trace of a star on a science slit, which is used to measure atmospheric seeing and throughput, as well as to account for any drift in telescope pointing by shifting individual frames to match the location of the star spectrum, which is particularly important when targeting the same mask for long periods without realigning due to flexure of the system \citep{Hutchison2020}.
The program weights each exposure according to the seeing and throughput values before coadding the frames and applying a telluric correction and flux calibration. 
Combined with the weighting of individual frames, this method returns 2D spectra with increases in SNR of $\sim5\%$ relative to the MOSFIRE DRP.
The resultant 2D spectrum was collapsed in a narrow range around either the strongest emission line or along the entire wavelength axis for continuum sources.
This was fit with a Gaussian which was subsequently used for weighting the 1D spectral extraction \citep[optimal extraction;][]{Horne1986}.

A modified version of the \texttt{spec2D} pipeline \citep{Cooper2012} was used to reduce C3VO DEIMOS spectra.
This program performs wavelength fitting, background sky subtraction, and 1D spectral extraction. Additional modifications are detailed in \citet{Lemaux2019} and include improvements in interpolation over the DEIMOS chip gap, throughput correction, and wavelength solution.

\subsubsection{Spectroscopic Redshift Determination}

The MOSFIRE spectra were visually inspected to identify galaxies with emission lines.
At the redshifts considered in this work, emission lines from \Hbeta\ and the \mbox{\OIII$\lambda\lambda$4959,5007} doublet are the most obvious features and when seen, a model consisting of three Gaussians is used to obtain a spectroscopic redshift.
This model has four parameters - the redshift, the amplitudes of \Hbeta\ and \OIIIfive, and the width of the lines.
The level of stellar continuum is assumed to be constant over the spectral range of the observations, and is taken to be the weighted average flux outside of regions with emission lines.
The amplitude of \OIIIfour\ is fixed to 30\% that of \OIIIfive\ \citep[\eg][]{Schreiber2018b}.
In this work we are not concerned with the quantification of line velocity widths, velocity offsets, or any broad line components beyond obtaining a redshift as pertains to MOSFIRE spectra.
A small number of galaxies at lower redshifts, which had \Halpha, \NII, and \SII\ in the wavelengths probed by the MOSFIRE observations were fit with a similar multi-Gaussian model.
Each spectroscopic observation is additionally given a quality flag in the style of the zCOSMOS, VUDS, and DEIMOS10k surveys described above which denotes the confidence in the assigned spectroscopic redshift \citep[see \eg][for more details]{LeFevre2015, Lemaux2022}.
Additionally, emission spatially offset from targeted galaxies that serendipitously fell in MOSFIRE slits (or in DEIMOS slits, see below) was re-extracted and the same redshift measurement and quality assessment process was performed on such detections.

The DEIMOS spectra were interactively assigned spectroscopic redshifts by using a modified version of the \texttt{zspec} environment \citep{Newman2013} as described in \citet{Lemaux2022}.
The modified version of the \texttt{zspec} software incorporates empirical high-redshift galaxy templates from the VUDS and VVDS \citep{LeFevre2004, LeFevre2013} surveys as well as high-resolution empirical Ly$\alpha$ templates from \citet{Lemaux2009}.
The quality flagging for these data follow a DEEP2 style flag \citep{Newman2013}.
Similar to the VUDS flagging system, flags of 3 and 4 represent high confidence redshifts ($>95\%$ accurate).
For example, the observation of a single emission line skewed redward indicative of Ly$\alpha$ would result in flag 3, and the observation of multiple spectral features would result in flag 4.
Flags of 1 and 2 indicate low confidence redshifts.
For consistency with the MOSFIRE data and redshifts from the field surveys, C3VO-DEIMOS objects with a quality flag of 2 are changed to a quality flag of 1 when creating the master spectroscopic catalog (next section).
All spectra were independently inspected and flagged by two of the authors (ES and BCL) and any objects receiving disparate redshifts/flags were reconciled via a re-inspection of the spectra and associated photometry (when available).

%----------------------------------------
\begin{figure*}
	\includegraphics[width=\textwidth, trim=0in 5.5in 0.5in 0in]{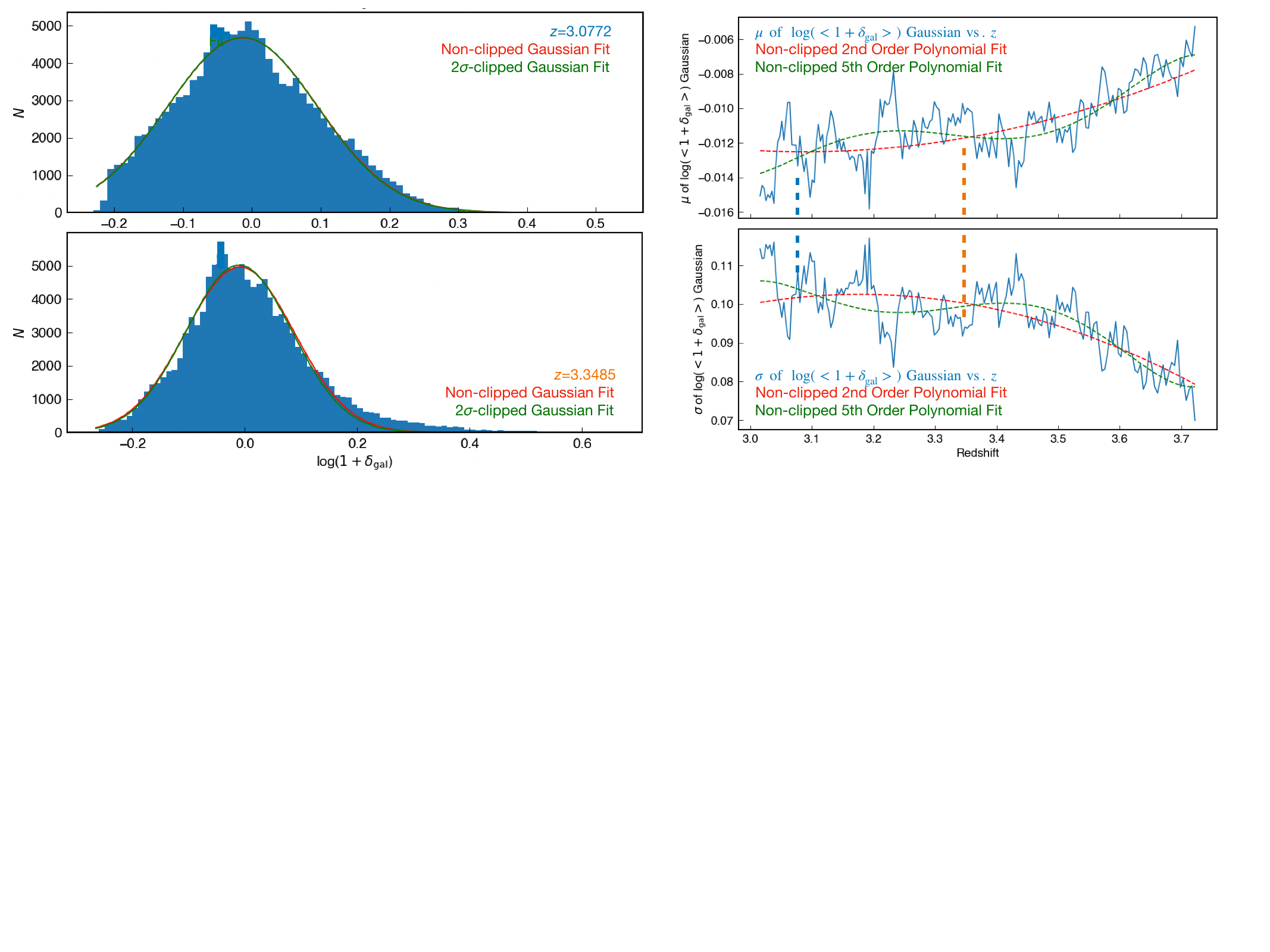}
    \caption{The process for determining the spread in voxel overdensity values for a given redshift slice. \textbf{Left:} Histograms of all voxel overdensity values (median set equal to zero) in redshift slices of $z=3.0772$ (top) and $z=3.3485$ (bottom). A Gaussian model is fit to the distribution at each slice, which produces a fit mean ($\mu$) and standard deviation ($\sigma$).
\textbf{Right:} The plot of the fit $\mu$ (top) and $\sigma$ (bottom) values for each slice of the VMC map considered in this work. A polynomial is then fit to this distribution to smooth out the effects of any large scale structures on these values, and these polynomial values are used to determine the overdensity significance ($\sigma_\delta$) of a given voxel. Herein we use the value of a $5^{\rm th}$-order polynomial (green), though using a $2^{\rm nd}$ order polynomial (red) for example does not effect the results. The redshifts of the two slices shown on the left are given by dashed vertical lines.}
    \label{fig:method}
\end{figure*}
%----------------------------------------

\subsection{Catalog Matching and Remodeling SEDs}

While target selection for the various spectroscopic surveys used in this work came from different parent photometric catalogs, we compare the spectroscopic redshifts, sky coordinates, and $i$- and $K$-band magnitudes from the spectroscopic survey catalogs to the photometric redshifts, coordinates and magnitudes in the same bandpasses from the COSMOS2020 Classic photometric catalog to find the best match.
For galaxies targeted in multiple surveys, only the highest quality flag entry was kept.
See Appendix~\ref{App:Matching} for more details regarding the matching process.
Of the 40008 final combined spectroscopic catalog entries, 37771 (94.4\%) have a match in the COSMOS2020 Classic catalog.
Additionally, 26676 (66.7\%) spectroscopic targets have moderate-high quality spectroscopic redshift measurements $0<z<7$ (quality flags X2, X3, X4, or X9).

With confirmed redshifts of these galaxies, it is necessary to remodel their physical properties, which will change from those given in COSMOS2020 unless the photometric redshift was identical to the spectroscopic redshift.
While there are many programs used to model galaxy spectral energy distributions (SEDs), in this work we use \texttt{LePhare} \citep{Arnouts1999, Ilbert2006}, as this program is used in the COSMOS2020 catalogs.
We remodel all spectroscopically confirmed galaxies using the same setup as in COSMOS2020 with the redshift now fixed to the spectroscopic redshift.
Detailed descriptions of these choices can be found in the catalog publications above, as well as \citet{Ilbert2009}.

As will be described shortly, we also perform a Monte Carlo resampling of the \pz\ for all galaxies without a high confidence spectroscopic redshift.
For each such galaxy we run \texttt{LePhare} with the redshift fixed to values from $3.0<z<3.7$ (the redshift range considered in this work) with $\Delta z = 0.05$.
The results from this grid are then interpolated to the redshift of a galaxy determined in an individual MC iteration.
We found that the differences in the fit stellar mass and SFR between adjacent redshift runs for a given galaxy differ by $>0.1$ dex in less than 0.3\% and 2.3\% of cases, respectively, more than sufficient for the studies herein, which validates this interpolation.

\section{Analysis}\label{Sec:Analysis}

\subsection{Voronoi Tesselation Monte Carlo Mapping}\label{Sec:VMC}

We use the Voronoi Tesselation Monte Carlo (VMC) mapping technique to determine galaxy environmental density, a method which has been extensively tested and used to find overdense structures previously \citep{Lemaux2017, Tomczak2017, Cucciati2018, Lemaux2018, Hung2020, Shen2021, Lemaux2022, Forrest2023}.
This method uses a combination of spectroscopic redshifts and photometric redshift probability distributions to statistically determine density in three-dimensional space.

\subsubsection{Density maps}

Voronoi cells are generated by drawing boundaries that are equidistant from the two nearest galaxies in projected space.
This strategy cannot be effectively extended to the redshift dimension due to the redshift uncertainties of objects in photometric catalogs as well as the uncertain contributions of peculiar velocities for spectroscopically confirmed galaxies.
While high spectroscopic completeness could minimize effects of the former, the spectroscopy in this work is too sparse to do so.
As a result of these uncertainties along the line of sight, the volume of interest is divided into redshift slices and only galaxies within such a slice are considered.
The slice width is determined from a combination of photometric redshift uncertainties and a consideration of overdensity sizes.
Structures can be missed either if slices are too narrow - when associated galaxies are not grouped together - or if slices are excessively wide, which results in a decrease in overdensity signal.
Similar to previous work, slices of 7.5 pMpc in depth are used ($\delta z\sim0.036$ at $z=3.35$) with an oversampling factor of $10\times$, that is the distance between central redshifts in adjacent slices is 0.75 pMpc.

We perform 100 Monte Carlo (MC) iterations, in which the \pz\ of every photometric galaxy is resampled to minimize the effect of photometric redshift uncertainties.
A statistical treatment of spectroscopic redshifts is also included, in which the quality flag of the spectroscopic redshift determines how often the spectral redshift is used as compared to a draw from that galaxy's \pz\ \citep[see Appendix B of ][]{Lemaux2022}.
In practice, the spectroscopically targeted galaxies with redshift quality flags of 3 or 4 have their spectroscopic redshift used in $\sim$99.4\% of all iterations (high confidence), galaxies with redshift quality flags 2 or 9 have their spectroscopic redshift used in $\sim$70.0\% of all iterations, and galaxies with redshift quality flags 0 or 1 do not have their spectroscopic redshifts used at all.
In such iterations where the spectroscopic redshift is not used, the redshift of a galaxy for that iteration is determined via a draw from the galaxy's \pz.

For each of the 100 iterations, Voronoi tesselation is performed on all objects with magnitude $\left[ 3.6 \right] <24.8$ that fall into a given redshift slice for a given iteration.
The results for each iteration are then regridded onto a regular grid of size $75\times75$ pkpc, with the median density at a grid point in all iterations assigned to each voxel, \dv.
Figure 3 of \citet{Tomczak2017} provides a visualization of this process.

\subsubsection{Overdensity maps}

The typical density of the Universe evolves with time, and the measurement of the average density from the VMC method is subject to differences in data quality, detection bands, magnitude cuts, etc.
As a result we are more concerned with the relative overdensity of a galaxy's environment rather than a pure density value.
This overdensity value is calculated by normalizing each density in a redshift slice by the median density of all voxels in the map at the same redshift, \mbox{\odv = log(1+\dv/ \mdv)}.
Additionally, given the extended nature of high-redshift protoclusters, the possibility exists that the entire field may be over- or underdense in a particular redshift slice, which could bias our overdensity calculations.
To account for this, we fit a Gaussian to the distribution of \odv\ values in each redshift slice, to obtain the average and standard deviation of the overdensity values therein (left panels of Figure~\ref{fig:method}).
A $5^{\rm th}$-order polynomial is then fit to the average overdensities as a function of redshift, removing any effects of field-filling over- or underdensities and obtaining a smoothed overdensity distribution, \odf\ (right panels of Figure~\ref{fig:method}).
We note that the differences between these two measures of overdensity, \odv\ and \odf, are on the order of several parts in one thousand, which is $\sim10\times$ smaller than the spread of values at a given redshift, and thus in general does not have a significant effect on our results.

Finally, the overdensity of a galaxy \od\ is simply the fit overdensity value \odf\ of the voxel which contains the galaxy's three-dimensional position in space in a given iteration.
We also calculate the number of standard deviations of a galaxy's overdensity value, $\sigma_\delta$, above or below the fit median using the fit standard deviation value at the redshift of interest.
This value will be the primary overdensity metric used in this work.

%----------------------------------------
\begin{figure}
	\includegraphics[width=0.45\textwidth, trim=0in 2.in 5in 0in]{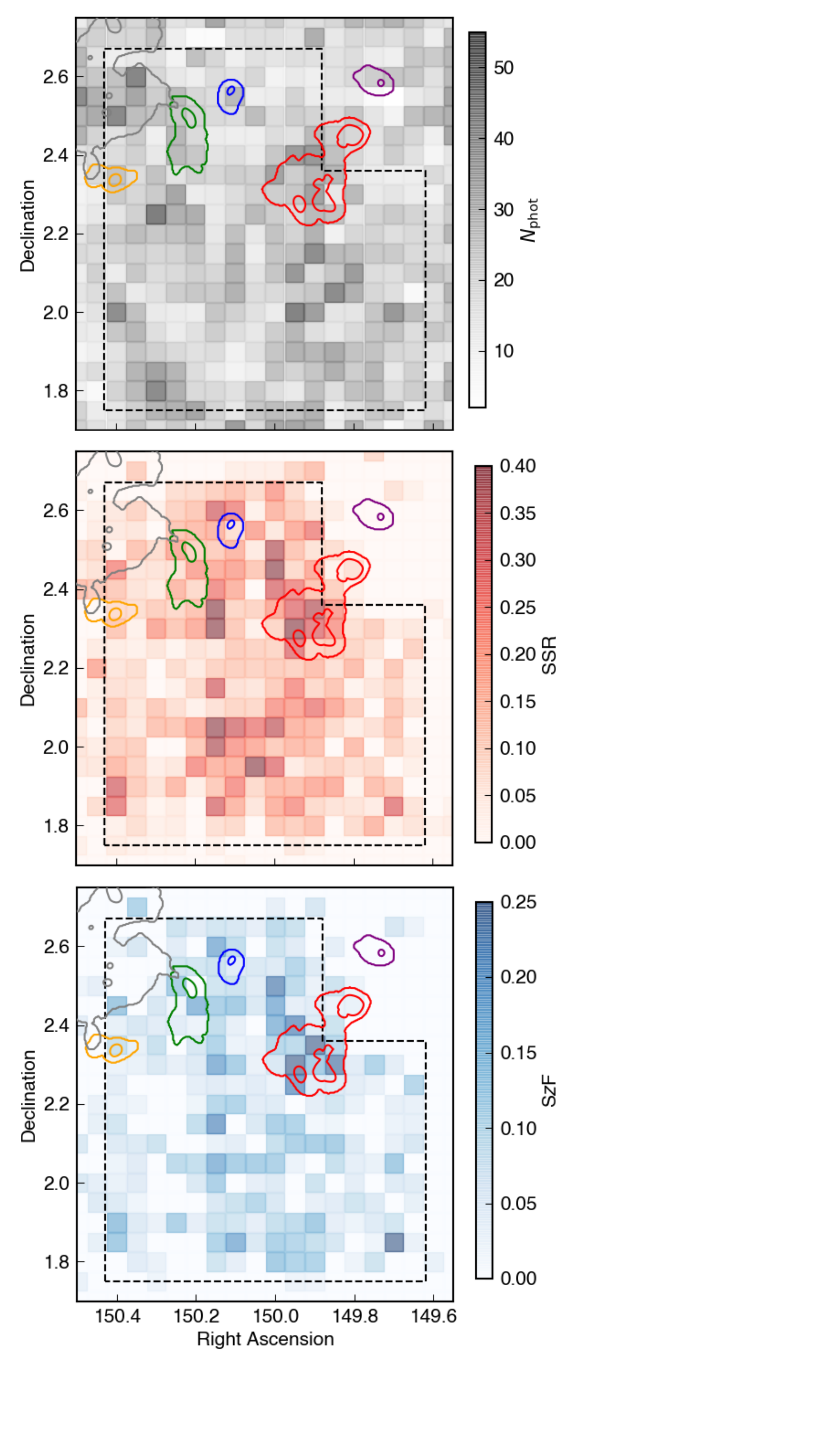}
    \caption{The relative numbers of photometric and spectroscopic galaxies in COSMOS in 3\arcmin$\times$3\arcmin\ bins across $3.0<z<3.7$ near the VUDS footprint (black dashed outline). \textbf{Top:} The number of galaxies from the COSMOS2020 catalog in 3\arcmin$\times$3\arcmin\ bins with photometric redshifts $3.0<z<3.7$. The structures associated with \PSC\ \citep{Forrest2023} are shown as contours representing $2\sigma$ and $5\sigma$ overdensities collapsed over $3.0<z<3.7$. \textbf{Middle:} Similar to the top panel, but showing the spectroscopic sampling rate (SSR). \textbf{Bottom:} Similar to the top panel, but showing the spectroscopic redshift fraction (SzF).}
    \label{fig:spat_all}
\end{figure}
%----------------------------------------

\subsection{Structure Identification}\label{Sec:Struct}

As described in \citet{Cucciati2018, Shen2021}, we identify overdense structures by finding all contiguous voxels with $\sigma_\delta>2$ ($>5$ for peak regions).
The volume enclosed within these envelopes can then be converted to a total mass based on the average comoving matter density, average overdensity of voxels within the envelope, and the galaxy sampling bias.
Following this process, a set of coeval structures with enclosed masses $\log(M_{\rm tot}/$M$_\odot)\gtrsim14$ in close proximity at $3.20<z<3.45$ were identified in \citet{Forrest2023}.
While that work focused on characterization of the structures themselves, in this work we analyze the galaxy populations therein.

In many studies of overdense structures, members are defined as being within some spherical radius of a central point in the structure.
However at early epochs such as this, protocluster systems are not necessarily spherically symmetric in nature, and the VMC maps allow for accounting and inclusion of some of these asymmetries.
This lack of symmetry also means that an accurate determination of the extent and overdensity of these structures is critical to drawing accurate conclusions about the effects such environments have on their component galaxies.

\subsubsection{Effects of Uneven Spectroscopic Sampling}

To this end, analyzing the dependence of overdensity significance on spectroscopic completeness is a critical test.
Using the above VMC method, a real overdensity with all members spectroscopically confirmed will have a stronger signal (higher $\sigma_\delta$) than the same overdensity with no spectroscopically confirmed members as the latter will have the total signal spread out in redshift space due to the width of the \pz\ for member galaxies.
Alternatively, if a region has little to no spectroscopically confirmed galaxies, it is possible that the VMC data may identify an overdensity when none is truly there.
Such a situation occurs when the \pz\ distributions of many galaxies in a small projected area, potentially including background and foreground objects, overlap significantly in redshift space.

We attempt to quantify these effects by considering galaxies with $\left[ 3.6 \right] <24.8$ and analyzing the spatial variance of: 1) the spectroscopic sampling rate (\specsamp) - the number of galaxies photometrically within a given volume which have been spectroscopically targeted divided by the total number of galaxies with a photometric redshift within that same volume, and 2) the spectroscopic redshift fraction (\szf), which we define here as the number of galaxies which have a high confidence spectroscopic redshift within some volume divided by the number of galaxies with a photometric redshift within that same volume.
We calculate these fractions in spatial bins of $3\arcmin \times 3\arcmin$ over the full $3.0<z<3.7$ redshift range considered for \PSC\ and the associated field.
This is shown in Figure~\ref{fig:spat_all} along with the number of objects with photometric redshifts in each bin.

As can be seen in the bottom two panels of Figure~\ref{fig:spat_all}, the \specsamp\ and \szf\ both fall off significantly outside of the VUDS footprint (dashed line) while the number of photometric targets stays roughly constant across the same boundary (top panel).
Overdensities identified in regions with low \szf\ have an increased probability of being false detections as the signal may be dominated by broad photometric redshift probability distributions, \eg\ S6 from \cite{Forrest2023}, shown by the very extended gray contour in the NE.
However, some regions with low \szf\ may still be real if there are significant spectroscopic confirmations, \eg\ the extension of S1 from \cite{Forrest2023} (red contours) outside of the VUDS footprint.
This particular region has 14 confirmed spectroscopic members from targeted follow-up with MAGAZ3NE \citep{McConachie2022}.
Clearly, further spectroscopic observations of potential overdensities in such regions can resolve this issue.

\begin{figure}
	\includegraphics[width=0.5\textwidth, trim=0in 0in 0in 0in]{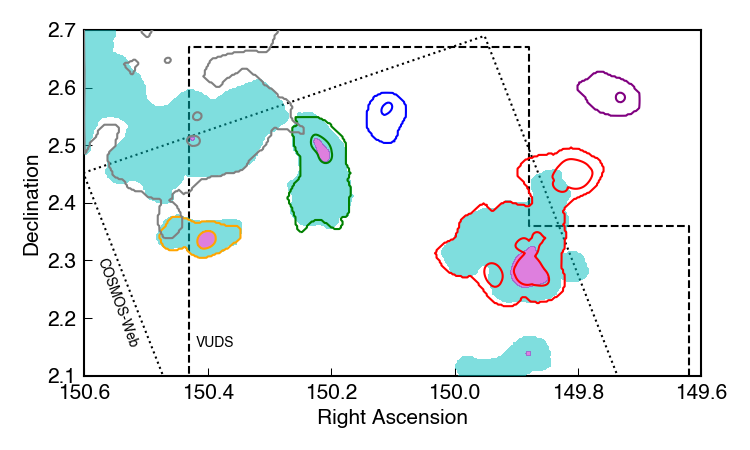}
    \caption{The recovery of overdensities in the field of interest with targeted follow-up spectroscopy (empty colored contours at +2$\sigma$ and +5$\sigma$) and without targeted follow-up spectroscopy (filled cyan and magenta contours at +2$\sigma$ and +5$\sigma$). The former set are as presented in \citet{Forrest2023}.}
    \label{fig:blind_check}
\end{figure}

The possibility also exists that overdensities identified in regions with high SzF and SSR, while real, are less overdense in reality than the VMC maps would suggest.
To ensure this effect is not influencing our results significantly, we rerun the VMC mapping analysis ignoring the spectroscopic redshifts from C3VO and MAGAZ3NE.
Removing this spectroscopy from our sample limits the effects of intentional spectroscopic targeting of overdense regions as all other surveys targeted galaxies across the COSMOS field as a whole.
Rerunning the VMC mapping and overdensity detection in this manner produces the results shown in
Figure~\ref{fig:blind_check}, where new structure overdensity contours of +2$\sigma$ and +5$\sigma$ are shown as filled cyan and magenta regions, respectively, and the original \PSC\ overdensities as reported in \cite{Forrest2023} are shown as open colored contours.

Using the same structure identification routine, the centers of commonly identified structures differ by medians of 78\arcsec\ (0.58 pMpc) in projection and 0.002 in redshift.
However, this can be split into those structures mostly within the VUDS footprint (S2 - orange contour on Fig.~\ref{fig:blind_check} - and S3 in green) with projected offsets of 8.1\arcsec\ (0.06 pMpc) and 22\arcsec\ (0.16 pMpc), and those which extend outside (S1 in red and S6 in gray) with projected offsets of 133\arcsec\ (0.99 pMpc) and 160\arcsec\ (1.2 pMpc).
For the former two structures the two maps result in volume differences of 8.0\% and 0.7\% and mass differences of 0.04~dex and 0.01~dex, in remarkable agreement.
The overdensity-weighted central three-dimensional locations for commonly identified +5$\sigma$ peaks differ in the two cases by similar amounts, with median differences of 18\arcsec\ (0.13 pMpc) in projection and 0.002 in redshift.
Noting that the typical systematic uncertainty for determining the stellar mass of a galaxy at similar epochs is estimated to be $\sim0.2$~dex \citep{Mobasher2015, Leja2019}, the characterization of these structures' masses and positions is robust to the effects of targeted spectroscopic followup at the levels obtained.
Regions extending beyond the VUDS outline without significant additional spectroscopy may be difficult to characterize accurately without additional data however.
On the whole, we conclude that the targeted spectroscopic followup has not significantly biased our characterization of the \PSC\ system, and for the remainder of this work consider the structure extends as in \citet{Forrest2023}.

%----------------------------------------
%----------------------------------------

\subsection{The Stellar Mass Function}

In order to investigate environmental effects on galaxy evolution, we construct stellar mass functions (SMFs) based on the overdensities derived from the VMC maps.
Similar to the generation of the VMC maps, we run 100 realizations drawing from the \pz\ of galaxies without reliable spectroscopic redshifts and determine the significance of the overdensity in which each galaxy resides, $\sigma_\delta$, from the VMC map at the galaxy's redshift in the realization.
All galaxies are subsequently refit using \texttt{LePhare} with the same methodology as when refitting the spectroscopically confirmed galaxies, but instead using the redshifts drawn from the \pz\ and the interpolated fits with $\delta z = 0.05$.
The resultant galaxy stellar masses are then used to construct SMFs.
The volume associated with a SMF is the volume of the VMC map within the associated redshift and $\sigma_\delta$ cuts.
 
For each realization, SMFs are constructed over \mbox{$3.20<z<3.45$} from galaxies in bins of $\sigma_\delta$: $-5<\sigma_\delta<2$ (field), $2<\sigma_\delta<3$, $3<\sigma_\delta<4$, and $\sigma_\delta>4$.
There are insufficient galaxies at $\sigma_\delta>5$ to recover a SMF in this regime with a large enough signal-to-noise ratio (SNR) to make significant conclusions.
Similarly, considering only galaxies within each structure of \PSC\ separately leads to small numbers of high-mass galaxies which make the uncertainties too large to justify drawing strong conclusions.
Consistent results are found from constructing a combined SMF for all galaxies in all regions of \PSC\ and from calculating the SMF based on overdensity values of \od\ as well.
We confirm that the field SMF is insensitive to other possible definitions, for example averaging galaxies with $-5<\sigma_\delta<2$ at $3.0<z<3.2$ and $3.5<z<3.7$
and that the field SMF is also in very good agreement with those over $3<$\zphot$<4$ based on the COSMOS2015 \citep{Davidzon2017} and COSMOS2020 \citep{Weaver2023a} catalogs.

%----------------------------------------
\begin{figure*}
	\includegraphics[width=\textwidth]{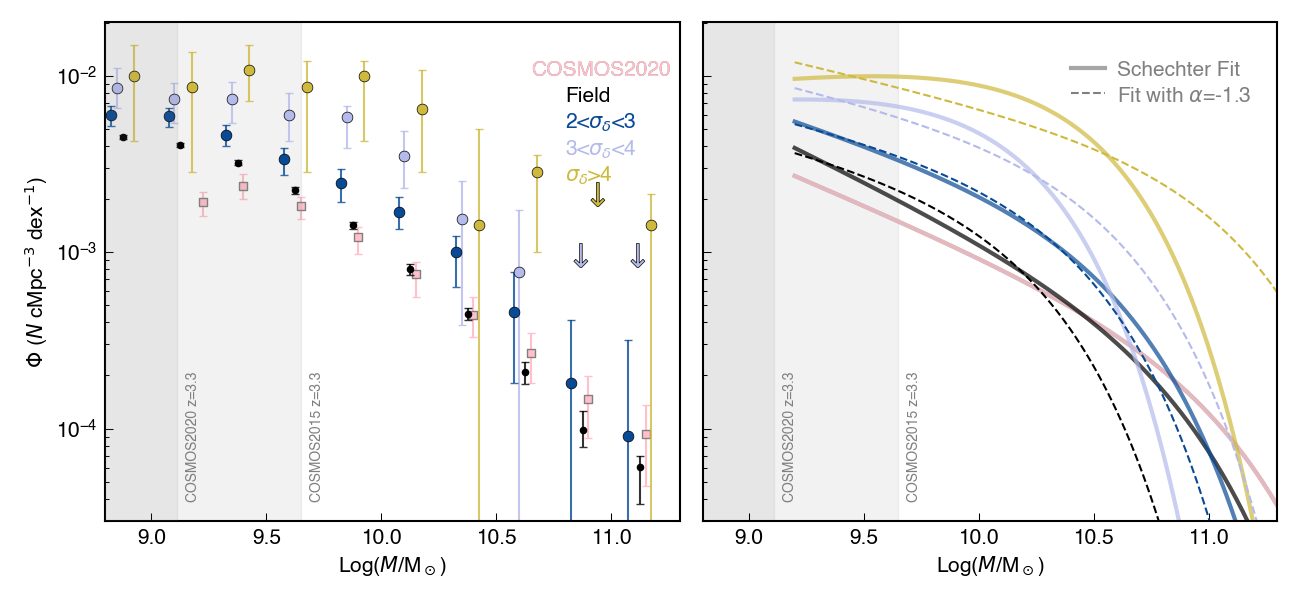}
    \caption{The stellar mass function in several bins of overdensity at $3.20<z<3.45$. \textbf{Left:} The SMF of the field from COSMOS2020 over $3.0<z<3.5$ \citep{Weaver2023a} is shown as a series of pink squares, while the field as measured in this work is shown as a series of black points. The SMFs of galaxies in bins of $2<\sigma_{\delta}<3$, $3<\sigma_{\delta}<4$, and $\sigma_{\delta}>4$ are shown as blue, periwinkle, and gold points, respectively. Error bars from this work represent the range of 16$^{th}$ to 84$^{th}$ percentiles from MC iterations added in quadrature to Poisson noise. Note that this does not include uncertainty due to cosmic variance, which is the dominant source of error in the COSMOS2020 analysis, and thus the error bars between the two works should not be compared. Masses below the stellar mass completeness limits at $z=3.3$ for the COSMOS2015 and COSMOS2020 catalogs are shaded. \textbf{Right:} The best-fit Schechter function to each measured SMF is shown as a solid line with the same color scheme. Another fit to each SMF performed fixing $\alpha$=-1.3 is shown as a dashed curve.}
    \label{fig:smf}
\end{figure*}
%----------------------------------------

\begin{figure}
	\includegraphics[width=0.5\textwidth]{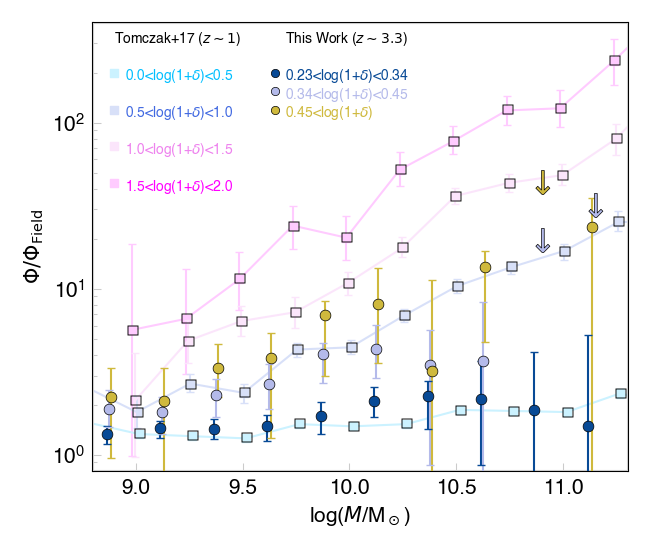}
    \caption{The ratio of the stellar mass function at different overdensities to that observed in the field. The gold, periwinkle, and blue points are as in Fig.~\ref{fig:smf} and represent galaxies in regions with overdensities of roughly $0.23<\log(1+\delta)<0.34$, $0.34<\log(1+\delta)<0.45$, and $\log(1+\delta)>0.45$ at $z\sim3.3$. Similar points from \citet{Tomczak2017} at $z\sim1$ are shown as squares for comparison. SMFs of similar overdensity seem to scale to the field in similar manners at both redshifts.}
    \label{fig:smf_rat}
\end{figure}

The median SMFs in each overdensity bin from all realizations are shown in Figure~\ref{fig:smf}, with error bars representing the 16$^{\rm th}$-84$^{\rm th}$ percentile range.
In addition to the vertical offset that is a result of the selected regions being overdense, a difference in the shape of the SMFs is also apparent and can be seen in Figure~\ref{fig:smf_rat}, in which the ratio of each SMF with that of the field is displayed.
In all overdense regions, there is a trend of increasingly elevated SMF relative to the field with increasing stellar mass, a pattern also seen in lower redshift cluster SMFs \citep{Tomczak2017, vanderBurg2020}, as well as in individual protoclusters at $z=2.16$ \citep{Shimakawa2018b} and $z=2.53$ \citep{Shimakawa2018a}.
This higher ratio of high-mass to low-mass galaxies in overdense regions is consistent with galaxies in protocluster environments undergoing, or having previously undergone, increased stellar mass build-up relative to field galaxies.

We compare our results to those from the $z\sim1$ ORELSE survey \citep{Tomczak2017}, which used a similar methodology for determination of overdensity.
The qualitative trends seen in the $z\sim1$ cluster sample are similar to those in this work, with the ratio of high-mass to low-mass galaxies increasing with increasing overdensity.
Additionally, as shown in Figure~\ref{fig:smf_rat}, when converting our SMFs from $\sigma_\delta$ to \od\ we find that the SMF for galaxies at similar overdensities appears the same at both $z\sim1$ and $z\sim3.3$.
The field of interest in this work and those in \citet{Tomczak2017} have different photometric bandpasses and depths, as well as different spectroscopic depths and SSR/SzF, which add uncertainty to such a direct comparison. 
However, this similarity of SMFs at similar overdensities at different redshifts would be consistent with the hypothesis that group environments, galaxy associations with lower total masses ($\sim 10^{13}$~M$_\odot$) than protoclusters, are host to evolutionary effects (\eg\ mergers) which `pre-process' galaxies before their eventual residence in cluster environments \citep{Zabludoff1996, McGee2009, DeLucia2012, Bahe2019, Reeves2021}.

\begin{figure}
	\includegraphics[width=0.5\textwidth, trim=0in 0.5in 2.5in 0in]{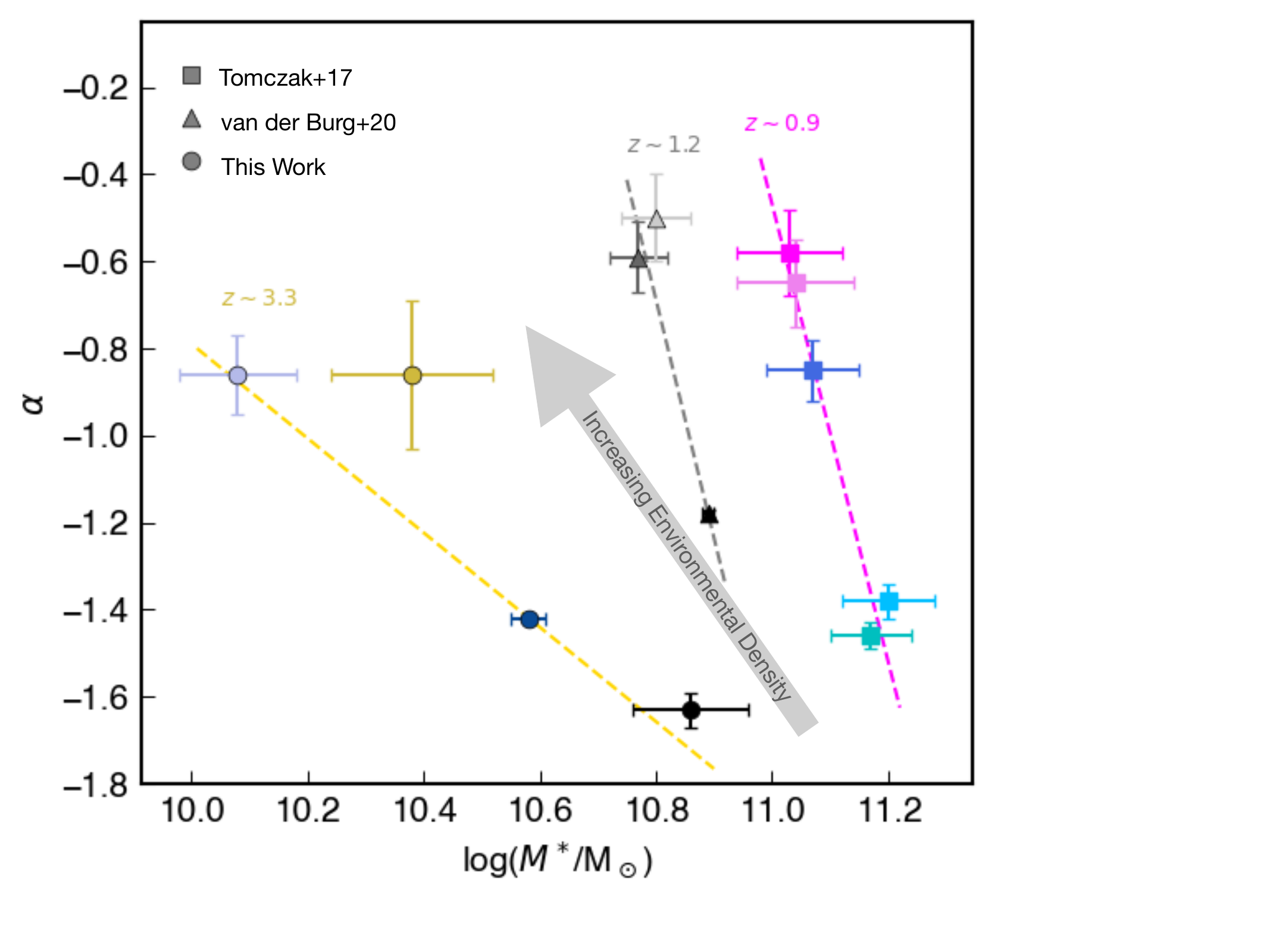}
    \caption{Comparison of the characteristic mass ($M^*$) and faint-end slope ($\alpha$) fit values from studies of the SMF in overdense environments. Values from this work are shown as circles of the same colors as in Fig.~\ref{fig:smf}, values from $z\sim1.2$ GOGREEN clusters \citet{vanderBurg2020} are grayscale triangles, and values from $z\sim0.9$ ORELSE clusters \citet{Tomczak2017} are cyan to magenta squares. A simple weighted linear fit for each dataset is shown in gold, gray, and magenta, respectively.}
    \label{fig:schpar}
\end{figure}

Additionally, we fit a single-Schechter function \citep{Schechter1976} to each SMF, first allowing the characteristic turnover mass ($M^{*}$), faint-end slope ($\alpha$), and normalization ($\phi^{*}$) parameters to vary, and then again fixing $\alpha=-1.3$, in rough agreement with previous results \citep[\eg][]{Marchesini2009, Muzzin2013, Tomczak2014}. 
The best-fit parameterizations are given in Table~\ref{tab:par}.
We compare the characteristic turnover mass and faint-end slope values to those from studies of $z\sim1$ cluster SMFs \citep{Tomczak2017,vanderBurg2020} shown in Figure~\ref{fig:schpar}.
In general, we find that $M^{*}$ decreases and $\alpha$ becomes shallower with increasing overdensity.
This is perhaps not so much due to the stellar mass of the average galaxy decreasing, but is instead a result of environmental processes increasing the number of galaxies around or just below the characteristic mass.
This increase may be due to increased merger rates in overdense environments \citep{Tomczak2017}, preferential enhancement of SFR in high mass galaxies in overdense environments, increased numbers of quiescent galaxies \citep{Nantais2016a, vanderBurg2018, vanderBurg2020}, or an underlying bias in galaxies in overdense environments relative to those in the field \citep{Ahad2024}.

Finally, we confirm that the observed difference in SMF shape is not due to preferential scattering of low mass galaxies with broader \pz\ out of the overdensity.
We use a toy model to resample redshifts, stellar masses, and overdensity membership of 1000 galaxies 1000 times in a mock region of sky, and then reconstruct SMFs (details in Appendix~\ref{A:SMFtest}).
No evidence for a mass-dependent bias is observed.

%----------------------------------------
%----------------------------------------

%----------------------------------------
%----------------------------------------

%----------------------------------------
\begin{table}
	\centering
	\caption{The parameters of Schechter function fits to the SMFs at varying densities. The best-fit values for varying $M*$, $\alpha$, and $\phi*$ are shown as are the results for fixing $\alpha=-1.3$.}
	\label{tab:par}
	\begin{tabular}{crrr}
		\hline
		Density bin & $\log(M^*/{\rm M}_\odot)$  & $\alpha$ & $\phi^*/(10^{-3} \rm{\ Mpc}^{-3})$ \\
		\hline
		$\sigma_\delta<2$ 	& 10.86$\pm$0.10	& -1.63$\pm$0.04	& 0.16$\pm$0.04	\\
						& 10.20$\pm$0.05	&=-1.3			& 0.87$\pm$0.07	\\
		$2<\sigma_\delta<3$	 & 10.58$\pm$0.03	& -1.42$\pm$0.01	& 0.66$\pm$0.04	\\
						& 10.40$\pm$0.03	&=-1.3			& 1.07$\pm$0.05	\\
		$3<\sigma_\delta<4$	 & 10.08$\pm$0.10	& -0.86$\pm$0.09	& 4.83$\pm$0.78	\\
						& 10.58$\pm$0.19	& =-1.3			& 1.49$\pm$0.33	\\
		$\sigma_\delta>4$	& 10.38$\pm$0.14	& -0.86$\pm$0.17	& 6.57$\pm$2.00	\\
						& 11.10$\pm$0.20	& =-1.3			& 1.40$\pm$0.35	\\
		\hline
	\end{tabular}
\end{table}
%----------------------------------------

\subsection{Quiescent Fractions}

The fraction of galaxies with SFRs below the main sequence of star formation (\ie\ quenched) at a given epoch is significantly increased in cluster environments relative to the coeval field out to $z\gtrsim1$ \citep{vanderBurg2013, Tomczak2017, vanderBurg2020}, though the majority of massive satellite galaxies at this epoch appear to have been quenched before cluster infall \citep{Baxter2022, Werner2022}.
High-redshift protocluster environments at $z\gtrsim2$ do not generally show these same quenched fractions, but some counterexamples appear to exist, at least with regards to high mass galaxies \citep{Chartab2020, Shi2021, Ito2023}, which has been seen in one of the peaks of \PSC\ \citep{McConachie2022}.
Analyzing how the SMFs of such quenched galaxies vary relative to the field environments can help uncover the mechanisms responsible for accelerated evolution of galaxies in overdense systems \citep[\eg][]{Tomczak2017,Papovich2018}.

For each of the 100 \pz\ sampling realizations, we calculate the number of quiescent and star-forming galaxies over $3.20<z<3.45$ in two ways.
First, we use the popular comparison of $(U-V)$ and $(V-J)$ rest-frame colors \citep[\eg][]{Williams2009, Muzzin2013a, Straatman2016} as modeled by \texttt{LePhare} and the wedge location specified for $2.0<z<3.5$ from \citet{Whitaker2011}.
We also fit the relation between SFR and stellar mass \citep[\eg][]{Daddi2007, Noeske2007, Salmon2015} by finding the median SFR in overlapping bins of width 0.2~dex with centers separated by 0.01~dex, smoothing via a first-order Savitzsky-Golay filter, and fitting with a quadratic polynomial.
Galaxies with SFRs more than 1~dex below this `star-forming main sequence' (SFMS) are then considered quiescent.
The resultant numbers of galaxies with \logM$>9.1$, the mass completeness of the COSMOS2020 Classic catalog at this redshift, are given in Table~\ref{tab:qf}.

\begin{figure*}
	\includegraphics[width=\textwidth, trim=0in 6.2in 0in 0in]{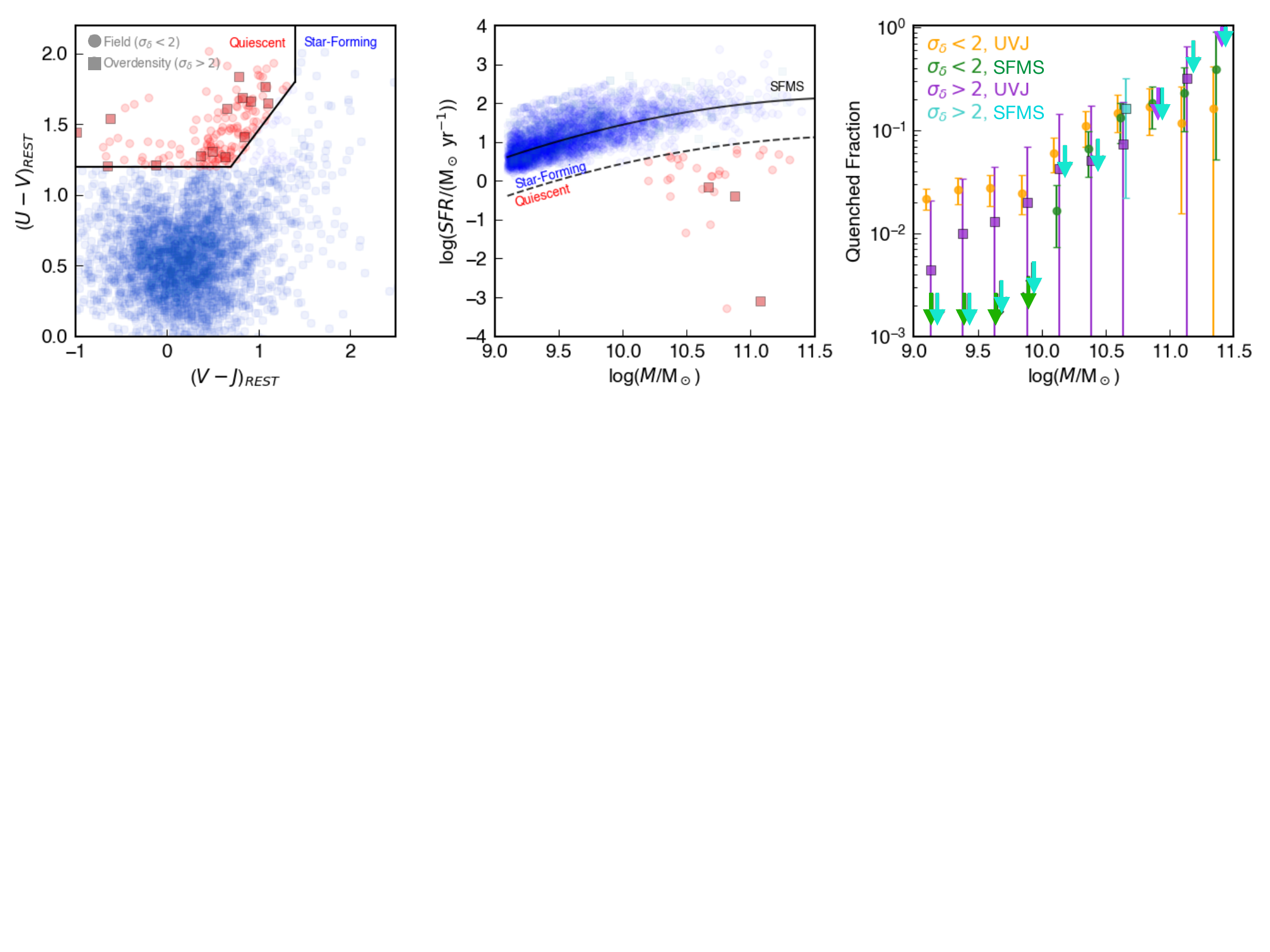}
    \caption{The identification of star-forming and quiescent galaxies at $3.20<z<3.45$.
      \textbf{Left:} The restframe $UVJ$ color-color plane for a single MC iteration. Red symbols above/left of the dividing wedge are considered quiescent, with squares indicating those galaxies in overdense environments and circles indicating galaxies in the field. Blue symbols below/right of the dividing wedge are classified as star-forming.
      \textbf{Center:} The SFR-stellar mass plane for a single MC iteration. The solid black curve is the median relation termed the star-forming main sequence (SFMS), while the dashed curve is 1~dex below the SFMS and is used to distinguish between star-forming and quiescent galaxies. The symbol colors and shapes remain the same.
      \textbf{Right:} The quenched fraction as a function of stellar mass averaged over all 100 MC iterations.
      Orange circles represent the field quenched fraction identified using the $UVJ$ selection.
      Green circles represent the field quenched fraction identified using the SFMS selection.
      Purple squares represent the overdensity quenched fraction identified using the $UVJ$ selection.
      Turquoise squares represent the overdensity quenched fraction identified using the SFMS selection.
      Downward-facing arrows indicate upper limits.
      While an increasing trend with stellar mass is seen, the sample sizes are insufficient to distinguish between the quenched fractions in field and overdense environments.
       }
    \label{fig:qf}
\end{figure*}

While the numbers of identified quiescent galaxies in the overdense regions are too small to construct SMFs from, we can compute quiescent fractions, which we plot in the right panel of Figure~\ref{fig:qf}, along with quiescent/star-forming selections for an example MC iteration in the left and middle panels.
No statistically significant differences between field and overdense environments are seen, though the uncertainties are significant.
The two methodologies also seem to be in general agreement, with the $UVJ$ color selection identifying more quiescent galaxies than the SFMS cut, particularly for galaxies with \logM~$<10$.

We do note however, that while the COSMOS2020 photometric catalog is complete down to \logM~$\sim9.1$ at $z\sim3.3$, most of the spectroscopy used for construction of the VMC maps is biased toward the detection of star-forming galaxies, which may artificially decrease the quenched fractions seen here.
The follow-up observations with MOSFIRE are less sensitive to this bias, but still no enhancement in quenched fraction relative to the field is seen in the overdense environments which were targeted with MOSFIRE.

%----------------------------------------
\begin{table}
	\centering
	\caption{The number of star-forming ($N_{SF}$) and quiescent galaxies ($N_Q$) and the associated quenched fractions (QF) in the field and overdensity samples across $3.20<z<3.45$ identified using $UVJ$ colors and the SFMS. Two sets of results are shown, one only considering galaxies with masses \logM~$>10.0$ and one considering galaxies with masses \logM~$>9.1$. }
	\label{tab:qf}
	\begin{tabular}{ccccc}
		\hline
		& $UVJ_{\rm M>9.1}$ & $UVJ_{\rm M>10}$ & SFMS$_{\rm M>9.1}$ & SFMS$_{\rm M>10}$\\
		\hline	
$N_{\rm SF, field}$	& 3593$^{+129}_{-137}$  		& 463.0$^{+37.0}_{-38.0}$  & 3688$^{+131}_{-136}$	 & 477.0$^{+39.0}_{-36.0}$ \\
$N_{\rm Q, field}$  	& 130.0$^{+40.0}_{-29.0}$  	& 50.0$^{+23.0}_{-14.0}$	  & 37.0$^{+17.0}_{-12.0}$	 & 37.0$^{+15.0}_{-12.0}$ \\
$N_{\rm SF, od}$    	& 398.0$^{+81.0}_{-71.1}$  	& 61.5$^{+24.0}_{-22.1}$    & 398.0$^{+88.6}_{-65.1}$	 & 59.0$^{+29.0}_{-18.9}$ \\
$N_{\rm Q, od}$     	& 7.5$^{+13.8}_{-7.1}$  		& 4.0$^{+6.3}_{-4.0}$  	  & 2.0$^{+5.4}_{-1.7}$		 & 2.0$^{+5.4}_{-1.7}$ \\
$QF_{\rm field}$    	& 3.49$^{+1.07}_{-0.78}$  	& 9.75$^{+4.48}_{-2.73}$    & 0.99$^{+0.46}_{-0.32}$	 & 7.20$^{+2.92}_{-2.33}$ \\
$QF_{\rm od}$       	& 1.85$^{+3.40}_{-1.74}$  	& 6.11$^{+9.66}_{-6.11}$  	  & 0.50$^{+1.34}_{-0.43}$	 & 3.28$^{+8.77}_{-2.84}$ \\
		\hline
	\end{tabular}
	\end{table}
%----------------------------------------

%----------------------------------------
%----------------------------------------

%----------------------------------------
\begin{figure*}
	\includegraphics[width=\textwidth]{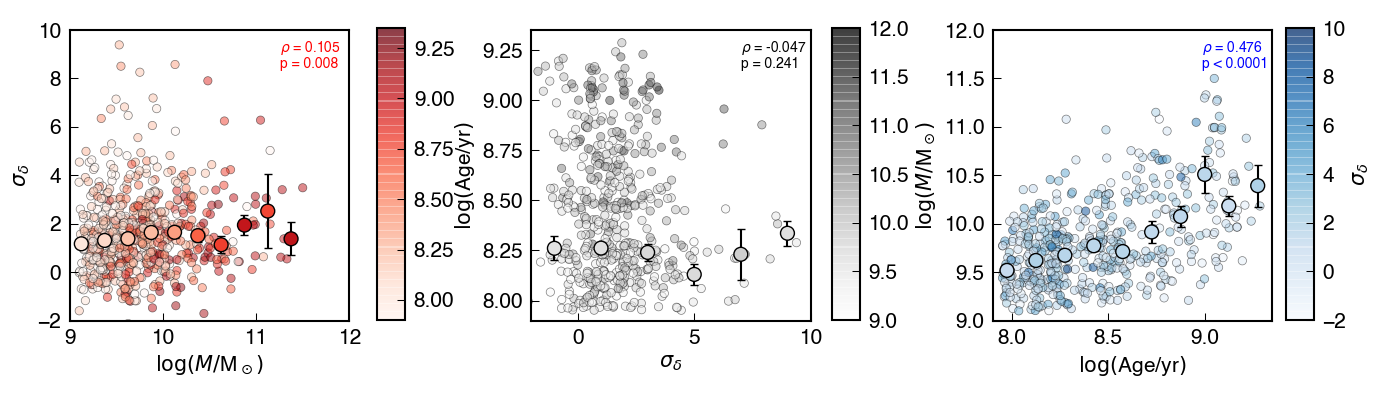}
    \caption{Comparison of stellar mass, stellar age, and environmental overdensity from SED fitting for spectroscopically confirmed galaxies at $3.0<z<3.7$ in COSMOS shown in three different projections. Median values are shown by the large circles, with the error bars representing the uncertainty on the calculated median. The Spearman $\rho$ coefficient and $p$-value are shown in each panel for the set of parameters on the ordinate and abcissa. }
    \label{fig:sedpar}
\end{figure*}
%----------------------------------------

\subsection{Relations between Stellar Mass, Age, and Overdensity}

Observed increases in the masses of galaxies residing in overdense environments relative to the field population could be caused by factors such as enhanced SFRs due to gas inflows or bursts of star formation as a result of increased merger rates.
Depending on when these effects occur, it is possible that the ages of galaxies of a given mass will have a dependence upon their environment as seen at lower redshifts \citep[\eg][]{Cooper2010a}, though these appear small at $z\sim1.3$ \citep{Webb2020}.
While in theory stellar ages can be discerned from spectral absorption features and the shape of the SED, the spectra of the vast majority of sources in this work do not have the requisite SNR to measure ages precisely, having spectral detections of emission lines only.
In an attempt to discern age differences between the field and structure populations, we fit the photometry of each spectroscopically-confirmed galaxy with the redshift fixed to the spectroscopic redshift using the \texttt{LePhare} \citep{Arnouts1999, Ilbert2006} and FAST++ \citep{Schreiber2018a} programs.
We then compare the stellar mass and age derived from these fits to the $\sigma_\delta$ of each galaxy and perform a Spearman correlation test (Figure~\ref{fig:sedpar}).

As seen in the SMF analysis, we find a weak ($\rho=0.105$), though significant ($p=0.008$) correlation between massive galaxies and overdense environments, similar to that found over $2<z<5$ in \citet{Lemaux2022}.
A commonly found correlation between stellar mass and age is also recovered ($\rho=0.476$, $p<0.0001$).
However, there is no significant trend for this sample between age and overdensity.
This could indicate that gas-rich mergers which trigger bursts of star formation are not responsible for the increased stellar masses of galaxies in these overdense environments, although some studies suggest that such bursts are not as pronounced at higher redshifts\citep{Shah2022}.
In this case, mergers could build high stellar mass galaxies without adding a significant younger stellar population.
Some other intrinsic difference between galaxies in overdense and field environments may also be responsible \citep{Ahad2024}.
It should be noted though that age determinations from SED fitting have significant uncertainties as well as degeneracies with stellar mass and dust extinction \citep[\eg][]{Mobasher2015}, and thus drawing strong conclusions from these tests is not supported.

\section{Conclusions}\label{Sec:Conc}

In this work we construct and compare stellar mass functions and quiescent fractions in several bins of environmental overdensity in the COSMOS field centered on the \PSC\ overdensity at $z\sim3.33$.
Such an analysis is only possible at these epochs with the extensive amounts of photometry and spectroscopy available in well-studied fields such as COSMOS.
These are used to build a three-dimensional density map to accurately estimate the environmental density field in which a galaxy resides.
We consider in this work galaxies above the approximate stellar mass completeness of the COSMOS2020 catalog at $z\sim3.3$, \mbox{\logM~$\sim9.1$}.

We observe distinct shapes of the stellar mass function between galaxies in overdense environments and galaxies in the field at $3.20<z<3.45$, with the densest regions having number densities $\sim6\times$ that of the field for galaxies with \mbox{\logM~$\sim10.0$}, compared to only $\sim3\times$ that of the field for galaxies with \mbox{\logM~$\sim9.5$}.
This distinction clearly indicates that the environment in which a galaxy resides begins to have an effect on its evolution prior to the galaxy entering a cluster environment.
The increased number of high mass galaxies in dense environments is suggestive of the long-theorized `pre-processing', in which the masses of such galaxies are enhanced in proto-cluster or even group systems via mergers and/or increased \textit{in situ} SFRs before infall into proper clusters, eventually resulting in the quenching of star formation in these overdense systems before quenching in field galaxies.
The quiescent fractions of galaxies in field and overdense environments do not differ, peaking at $\sim 20-30$\% for galaxies with \mbox{\logM~$\sim11.0$}.
This may suggest that while processes of mass enhancement in overdense environments have begun, processes which enhance the quenched fraction in these same environments have not yet had significant effects.
We note however, that the uncertainties on these measurements are large and that the spectroscopic surveys in this work are in general more sensitive to star forming populations.
More dedicated observations of quiescent candidates identified from photometry are required for stronger conclusions.

Finally, we compare the relationship between stellar mass, stellar age, and environmental overdensity of spectroscopically-confirmed galaxies.
We find statistically significant positive correlations between stellar mass and stellar age, as well as between stellar mass and environmental overdensity.
This again suggests that the overdense structure of the \PSC\ proto-supercluster may be affecting the evolution of member galaxies through enhanced stellar mass growth, but is inconclusive on the matter of enhanced quenching of member galaxies.
Further observations, particularly of the densest protocluster cores in \PSC\ are necessary to uncovering the extent and significance of these effects.

\section{Acknowledgements}

The authors wish to recognize and acknowledge the very significant cultural role and reverence that the summit of Maunakea has always had within the indigenous Hawaiian community.  We are most fortunate to have the opportunity to conduct observations from this mountain.
This work is also based on observations collected at the European Southern Observatory under ESO programmes
\mbox{175.A-0839}, % zCOSMOS
\mbox{179.A-2005}, and % UltraVISTA
\mbox{185.A-0791}, % VUDS
as well as work supported by the National Science Foundation under Grant No. 1908422.
GW gratefully acknowledges support from the National Science Foundation through grant AST-2205189 and from HST program number GO-16300.

This work has relied heavily upon code developed by other people, for which we are quite thankful.
\software{
Astropy \citep{Astropy2013,Astropy2018,Astropy2022},
FAST++ \citep{Schreiber2018a},
IPython \citep{Perez2007},
LePhare \citep{Arnouts1999, Ilbert2006},
Matplotlib \citep{Hunter2007},
NumPy \citep{Oliphant2006}.
}

%%%%%%%%%%%%%%%%%%%% REFERENCES %%%%%%%%%%%%%%%%%%

\bibliography{library}

%%%%%%%%%%%%%%%%% APPENDICES %%%%%%%%%%%%%%%%%%%%%

\appendix
\renewcommand\thefigure{\thesection.\arabic{figure}}

\section{Object Matching}\label{App:Matching}
\setcounter{figure}{0}

We use observations of galaxies from the zCOSMOS \citep{Lilly2007}, VUDS \citep{LeFevre2015}, and DEIMOS10k \citep{Hasinger2018} spectroscopic surveys, as well as observations taken with Keck/MOSFIRE as part of the C3VO \citep{Lemaux2022} and MAGAZ3NE \citep{Forrest2020b} surveys to construct a master spectroscopic catalog.
The targets in these samples are drawn from various observational catalogs in the COSMOS field and here we detail the process of matching these observations to galaxies in the COSMOS2020  photometric catalog \citep{Weaver2022}.

\subsection{Astrometric Correction}

For each spectroscopic survey, each spectroscopic entry ($s_i$) is matched to the nearest photometric catalog member ($p_{j0}$) in projected space.
The coordinates for spectroscopic entries are then updated based on the median positional offsets between the data sets, median($\Delta \alpha_{i,j0}$) and median($\Delta \delta_{i,j0}$), providing a first-order astrometric correction.
These median offsets are $\lesssim0.1''$ arcseconds in all cases.

\subsection{Positional Threshold}
\begin{figure*}[!hb]
	\includegraphics[width=\textwidth]{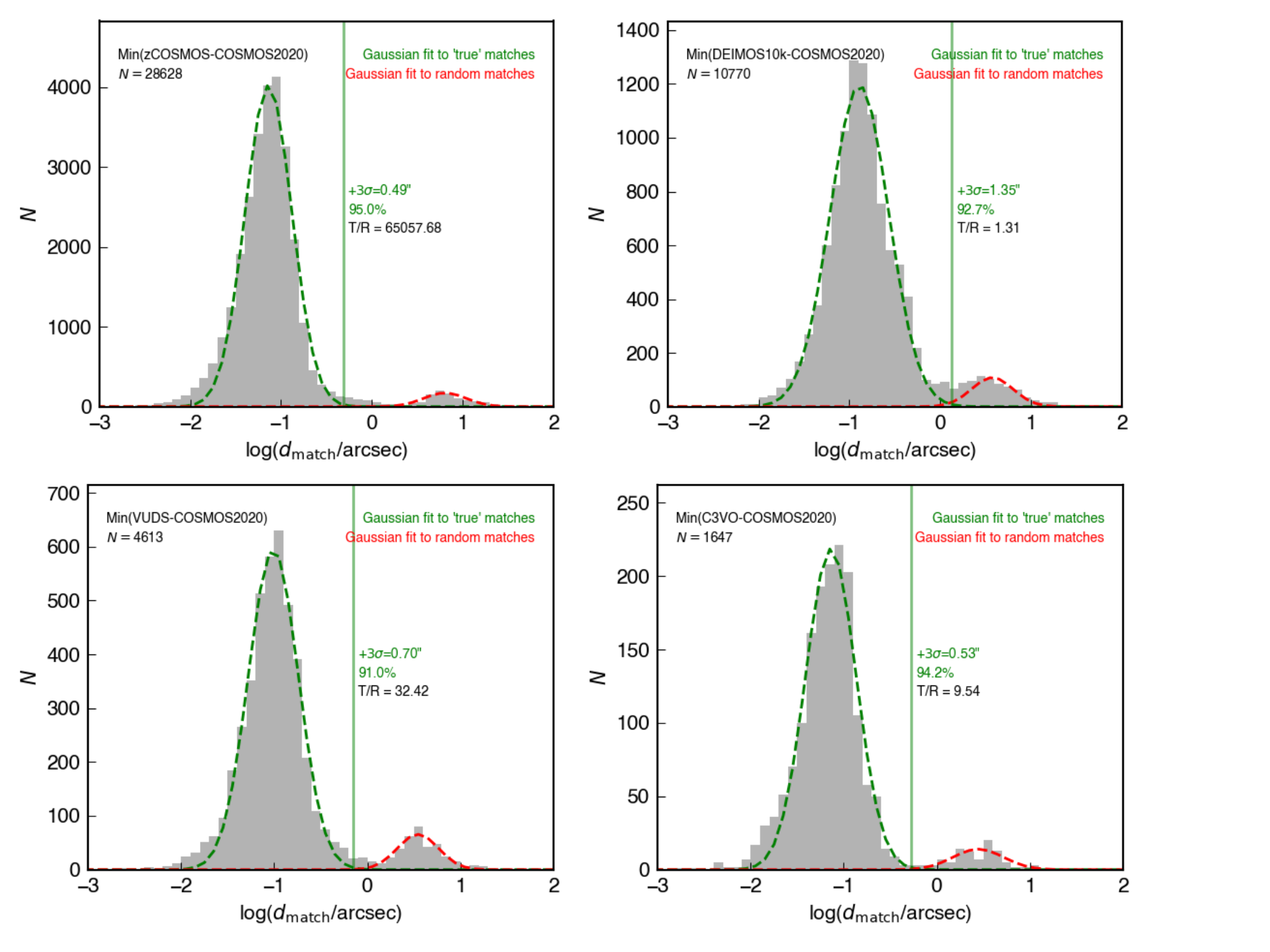}
    \caption{Determination of matching thresholds. The distribution of distances from each spectroscopic object to the nearest photometric catalog member after a bulk astrometric correction is plotted as a gray histogram. A two Gaussian model is fit, with the green representing `true' matches and the red representing random matches. A green vertical line shows the $+3\sigma$ value of the `true' Gaussian fit, with the associated separation, percentage of matches within that separation, and the ratio of amplitude of the two Gaussian components at that distance given. }
    \label{fig:match_dm}
\end{figure*}

Catalog matching is performed again to find the nearest photometric catalog member ($p_j$).
We then use the distribution of distances between matches to calculate our matching tolerance, considering that this distribution is composed of a combination of correct matches and random nearest matches in cases where the spectroscopic target is not in the photometric catalog.
These two components become clear by analyzing the histogram of logarithmic distance separations (Figure~\ref{fig:match_dm}).

We fit a Gaussian to each of these distributions and take the $3\sigma$ upper limit for the main peak (assumed to be correct matches) to be the distance threshold within which to search for a given spectroscopic survey.
Alternatively taking the $3\sigma$ limit from fitting a single Gaussian to the entire distribution or by choosing the distance at which the contribution of the two Gaussians are equal, while changing the number of galaxies matched, does not result in differences to the scientific conclusions of this work.
For objects in the VUDS, C3VO, zCOSMOS, and DEIMOS10k surveys, this search radius threshold is  0.70", 0.53", 0.49", and 1.35", respectively, which successfully matches 91.0\%, 94.2\%, 95.0\%, and 92.7\% of entries.

%----------------------------------------
%----------------------------------------

\subsection{Single Matches}
If there is a single photometric match to a spectroscopic object within the $3\sigma$ radius threshold of a survey, we take the two entries to be matches to the same object.

\subsection{Multiple Matches}
Many of the spectroscopic entries in the catalog have more than one photometric catalog member which satisfies both the distance and magnitude threshold cuts above.
In these cases, comparisons are also made between the spectroscopic redshift and the photometric redshifts of potential matches, as well as differences in their total $i$- and $K$-band magnitudes, resulting in a four-parameter comparison.
If no other spectroscopically-confirmed galaxies are nearby, the photometric member with the better weighted combination of values is determined to be the correct match.
If another spectroscopic entry is also nearby, the photometric and spectroscopic entries are paired based on closest positional match.

\subsection{No Matches}
There are 2237 spectroscopic entries in the catalog which have no photometric catalog members that satisfy the distance threshold cuts above.
These objects are retained in the final spectroscopic catalog and are not considered when generating the VMC maps or further analyses as they do not have the requisite photometry for performing SED fitting to determine stellar mass, SFR, etc.

\subsection{Duplicates}
At this point, each spectroscopic object is considered to be matched to the correct photometric catalog object.
As such, photometric catalog objects with multiple spectroscopic matches exist and are given a true multi-spec flag.
If these spectroscopic entries (which have a quality flag of 3, 4, or 9) have the same redshift, this is considered the spectroscopic redshift.
When discrepant spectroscopic redshifts with identical good quality flags exist, the instrument and survey are used to determine the correct redshift.
In order of priority, these are: C3VO, MAGAZ3NE, VUDS, DEIMOS-10k, zCOSMOS (updated catalog), zCOSMOS (original catalog).

\clearpage

\section{SMF Significance Testing}\label{A:SMFtest}
\setcounter{figure}{0}

It is possible that the observed difference in the SMF of members of the Elent\'ari protostructure and the coeval field could be partially or entirely due to differential effects related to large photometric redshift uncertainties.
More specifically, assuming that lower mass galaxies have broader \pz , they are more likely to scatter in and out of different environmental bins than high mass galaxies.
If both the field and overdensity SMF have the same intrinsic shape and are simply different in normalization, then due to the fairly small redshift extent of the higher density bins presented in this work, a broader \pz\ for lower mass galaxies relative to their more massive counterparts would potentially lead to a differential loss of such galaxies within the highest density bins as they will scatter out of the volumes associated with the higher density environments and into those associated with the lower density environments more frequently than higher mass galaxies.
Under such a scenario the SMF in higher density bins would then appear to have a lower ratio of low-mass to high-mass galaxies than the field.

We use a toy model to test whether such an effect can reproduce the observed results.
We start by assuming that the only difference in the field and overdensity SMFs is the normalization (factor of 10 difference) and that overdensities have a volume filling factor of $\sim3\%$ \citep{Chiang2017}, and then populate a mock region of sky with 1000 galaxies.
The probability of a random galaxy residing in the field/overdensity will then be 
\begin{eqnarray}
p_f &=& \frac{0.97\times1}{0.03\times10 + 0.97\times1}\times100\% = 76.4\% \\
p_o &=& \frac{0.03\times10}{0.03\times10 + 0.97\times1}\times100\% = 23.6\%
\end{eqnarray}

Field galaxies are distributed randomly across the mock projected area and assigned a random redshift in the range $3.0<z<3.7$.
Overdensity members are assigned positions based on random draw from a Gaussian in each dimension such that the $\pm2\sigma$ values encompass 3\% of the total volume.
The centers and widths of these Gaussians are $\mu_\alpha=\mu_\delta=0$, $\mu_z=3.325$, $\sigma_\alpha$=$\sigma_\delta$=0.14, $\sigma_z$=0.0625.
Each galaxy is then assigned a stellar mass drawing from the SMF.

With each galaxy assigned a `true' position, redshift, and stellar mass, we now assign redshift probability distributions based on uncertainties from the COSMOS2020 catalog.
Each galaxy is assigned a \pz\ which is an asymmetrical Gaussian.
The center of this Gaussian is drawn from a Gaussian characterized by the mean and standard deviation of the offset between spectroscopic and photometric redshifts at $3.0<z<3.7$ ($0.09\pm0.29$).
Similarly, the width of each side of the Gaussian is drawn from a Gaussian characterized by the mean and standard deviation of the $16^{\rm th}$ and $84^{\rm th}$ percentiles of the \pz\ of galaxies at similar stellar mass at $3.0<z<3.7$.

We then run 1000 Monte Carlo iterations, in each one randomly drawing a `measured' redshift from the \pz\ of each galaxy.
A `measured' stellar mass is determined from the difference between `measured' and `true' redshifts combined with an uncertainty of 0.2~dex \citep[\eg][]{Mobasher2015,Wang2023a}.
The overdensity or field membership of a galaxy is then determined based on the three dimensional position of the galaxy relative to the known overdensity extent.
A SMF is then constructed for the field and the overdense regions using the medians of all 1000 iterations with uncertainties given by the 16$^{th}$ and 84$^{th}$ percentile values.

\begin{figure*}[!hb]
	\includegraphics[width=\textwidth]{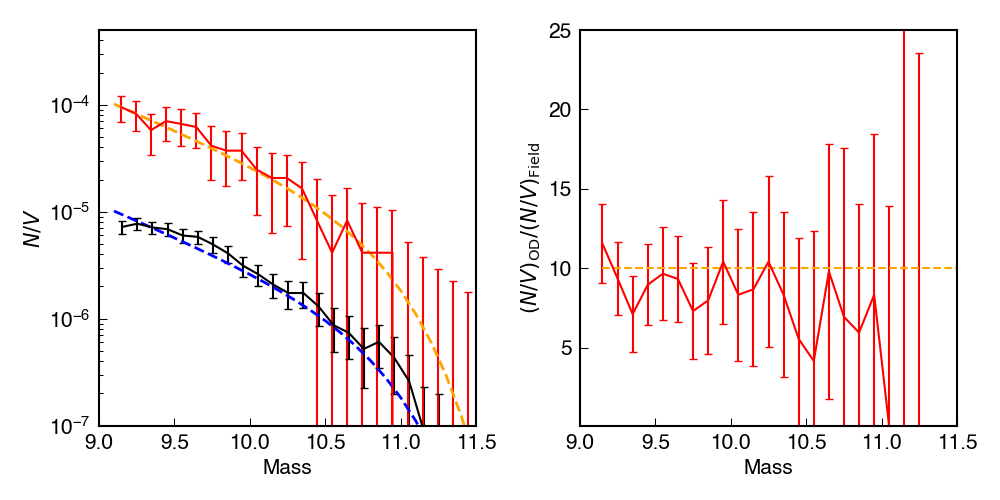}
    \caption{The result of an example run of the model to test for the effect of biases on the SMF ratios. \textbf{Left:} The generating SMFs for field and overdensity are shown as blue and orange dashed curves, respectively, and the recovered SMFs are plotted as red and black lines. \textbf{Right:} The ratio of the generating SMFs (orange dashed line) and recovered SMFs (red line). No bias toward higher SMF ratios at higher masses is observed.}
    \label{fig:SMF_test}
\end{figure*}

This process is run ten times to remove any bias due to assignment of galaxy locations.
In all runs, the ratio between the overdensity and field SMF shows no evidence of the increased high-mass to low-mass galaxy ratio seen the observations. 
The result of an example run is shown in Figure~\ref{fig:SMF_test}.
We note that this test does not include consideration of spectroscopic redshifts which would reduce the size of uncertainties further.

%----------------------------------------
%----------------------------------------

%%%%%%%%%%%%%%%%%%%%%%%%%%%%%%%%%%%%%%%%%%%%%%%%%%

\end{document}